\documentclass[12pt]{article}

\renewenvironment{abstract}
{\begin{quote}
\noindent \rule{\linewidth}{.5pt}\par{\bfseries \abstractname.}}
{\medskip\noindent \rule{\linewidth}{.5pt}
\end{quote}
}

\usepackage{bbm}
\usepackage[numbers]{natbib}
\usepackage[dvipsnames]{xcolor}

\usepackage[draft,authormarkuptext=name,authormarkup=superscript, authormarkupposition=right]{changes}

\usepackage{tcolorbox}
\definechangesauthor[name=HKSayka, color=magenta]{1}
\definechangesauthor[name=HKSayka, color=green]{2}
\definechangesauthor[name=HKSayka, color=gray]{6}
\definechangesauthor[name=HKSayka, color=orange]{7}
\definechangesauthor[name=HKSayka, color=blue]{8}
\definechangesauthor[name=HKHK, color=black]{9}
\definechangesauthor[name=SaykaHK, color=Periwinkle]{3}
\definechangesauthor[name=SaykaSayka, color=orange]{4}
\definechangesauthor[name=HKSaykaDone, color=gray]{5}

\usepackage{silence}
\ActivateWarningFilters[pdftoc]

\newcommand{\chapfnt}{\fontsize{16}{16}}
\newcommand{\secfnt}{\fontsize{16}{16}}
\newcommand{\ssecfnt}{\fontsize{16}{16}}

\makeatletter
\def\@makechapterhead#1{%
  \vspace*{50\p@}
  {\parindent \z@ \raggedright \normalfont
    \ifnum \c@secnumdepth >\m@ne
        \chapfnt\bfseries \@chapapp\space \thechapter
        \par\nobreak
        \vskip 20\p@
    \fi
    \interlinepenalty\@M
    \chapfnt \bfseries #1\par\nobreak
    \vskip 40\p@
  }}
\renewcommand\section{\@startsection {section}{1}{\z@}%
                                   {-3.5ex \@plus -1ex \@minus -.2ex}%
                                   {2.3ex \@plus.2ex}%
                                   {\normalfont\secfnt\bfseries}}
\renewcommand\subsection{\@startsection{subsection}{2}{\z@}%
                                     {-3.25ex\@plus -1ex \@minus -.2ex}%
                                     {1.5ex \@plus .2ex}%
                                     {\normalfont\ssecfnt\bfseries}}
\usepackage{geometry}
\usepackage{tabularx}

 \newlength{\dhatheight}
\newcommand{\plus}[1]{\hat{#1}}
\newcommand{\plusplus}[1]{\plus{\plus{#1}}}
\newcommand{\minus}[1]{\check{#1}}
\newcommand{\minusminus}[1]{\minus{\minus{#1}}}

\newcommand{\dt}{\cdot}

 \usepackage{setspace}

\usepackage{ulem}
\usepackage{color,soul}

\setul{0.5ex}{0.3ex}
\definecolor{Green}{rgb}{0,1,0}
\setulcolor{Green}

\definecolor{Red}{rgb}{1,0.0,0.0}
\setulcolor{Red}

\definecolor{Blue}{rgb}{0,0.0,1}
\setulcolor{Blue}

\usepackage[cal=boondox,scr=boondoxo]{mathalfa}
\DeclareFontEncoding{FML}{}{}
\DeclareFontSubstitution{FML}{fncmi}{m}{it}
\DeclareSymbolFont{fourierletters}{FML}{fncmi}{m}{it}
\SetSymbolFont{fourierletters}{normal}{FML}{fncmi}{m}{it}
\DeclareMathSymbol{\fP}{\mathalpha}{fourierletters}{`E}
\DeclareSymbolFont{Nperm}{OML}{cmss}{bx}{it}

\SetSymbolFont{Nperm}{bold}{OML}{cmss}{bx}{it}
\DeclareMathSymbol{\nw}{\mathalpha}{Nperm}{`w}
\newcommand{\unit}[1]{\ensuremath{\, \mathrm{#1}}}
\usepackage{pagecolor}

\usepackage{amssymb,amsmath,textcomp,gensymb,amsthm,mathtools}

\usepackage{lineno}

\usepackage{tabularx}
\numberwithin{equation}{section}

\usepackage[T1]{fontenc}
\usepackage{lmodern}
\usepackage{xcolor}

\usepackage{tikz}
\usepackage{cutwin}
\usepackage[framemethod=TikZ]{mdframed}
\usetikzlibrary{arrows.meta,
                positioning,
                quotes,shapes,arrows,fit,calc,automata
                }
\usetikzlibrary{decorations.markings}

\DeclareRobustCommand\full  {\tikz[baseline=-0.6ex]\draw[-stealth, line width=1pt] (0,0)--(0.6,0);}
\DeclareRobustCommand\dotted{\tikz[baseline=-0.6ex]\draw[-stealth, line width=1pt,dotted] (0,0)--(0.64,0);}
\DeclareRobustCommand\dashed{\tikz[baseline=-0.6ex]\draw[-stealth, line width=1pt,dashed] (0,0)--(0.64,0);}

\DeclareRobustCommand\opentriangle  {\tikz[baseline=-0.6ex]\draw[-{Triangle[open]}, line width=1pt] (0,0)--(0.6,0);}
\DeclareRobustCommand\openkite  {\tikz[baseline=-0.6ex]\draw[-{Kite[open]}, line width=1pt] (0,0)--(0.6,0);}

\usepackage{enumitem}
\usepackage{pifont}
\usepackage{epstopdf}
\usepackage{xr}
\usepackage{xstring}
\usepackage{float}
\usepackage[]{hyperref}
\hypersetup{
colorlinks = true,
citecolor = blue,
linkcolor = blue
}

\usepackage{xifthen}
\usepackage{xargs}

\newcommand{\physf}{\boldsymbol{\mathscr{f}}}
\newcommand{\physg}{\boldsymbol{\mathscr{g}}}

\newcommand{\physga}{\boldsymbol{\mathscr{g}}}
\newcommand{\physgb}{\boldsymbol{\mathscr{h}}}
\newcommand{\physe}{\boldsymbol{\mathscr{e}}}
\newcommand{\physx}{\boldsymbol{\mathscr{x}}}
\newcommand{\phys}[1]{\boldsymbol{\mathscr{#1}}}

\newcommand{\pr}[1]{\left( #1 \right)}
\newcommand{\ar}[1]{\left[ #1 \right]\,}
\newcommand{\ag}[1]{\left[ #1 \right]}

\newcommand{\fp}{\physf_{\varphi}}

\newcommand{\sgC}[4]{s^{\physg}_{\pr{(#1,#2),(#3,#4)}}}
\newcommand{\sgM}[2]{s^{\physg}_{\pr{#1,#2}}}
\newcommandx*{\sg}[3][1=,2=,3=,usedefault]{%
  s^{\physg}_{\ifthenelse{\isempty{#1}}{}{\ifthenelse{\isempty{#2}}{#1}{#1 \cdot \scriptscriptstyle{#2} \cdot \scriptscriptstyle{#3}}}}
}

\newcommandx*{\se}[5][1=,2=,3=,4=,5=,usedefault]{%
  s^{\physe}_{
  \ifthenelse{\isempty{#1}}
  {}
  {
  \ifthenelse{\isempty{#2}}
  {#1}
  {\ifthenelse{\isempty{#4}}
  {
  #1 \cdot \scriptscriptstyle{#2} \cdot \scriptscriptstyle{#3}
  }
  {
  #1 \cdot \scriptscriptstyle{#2} \cdot \scriptscriptstyle{#3}\cdot \scriptscriptstyle{#4} \cdot \scriptscriptstyle{#5}}
  }
  }
  }
}%
\newcommand{\darg}{x}

\renewcommandx*{\sf}[5][1=,2=,3=,4=,5=,usedefault]{%
  s^{\physf_{\varphi}}_{
  \ifthenelse{\isempty{#1}}
  {}
  {
  \ifthenelse{\isempty{#2}}
  {#1}
  {\ifthenelse{\isempty{#4}}
  {
  #1 \cdot \scriptscriptstyle{#2} \cdot \scriptscriptstyle{#3}
  }
  {
  #1 \cdot \scriptscriptstyle{#2} \cdot \scriptscriptstyle{#3}\cdot \scriptscriptstyle{#4} \cdot \scriptscriptstyle{#5}}
  }
  }
  }
}%

\newcommand{\cgC}[4]{c^{\physg}_{\pr{(#1,#2),(#3,#4)}}}
\newcommand{\cgM}[2]{c^{\physg}_{\pr{#1,#2}}}
\newcommandx*{\cg}[3][1=,2=,3=,usedefault]{%
  c^{\physg}_{\ifthenelse{\isempty{#1}}{}{\ifthenelse{\isempty{#2}}{#1}{#1 \cdot \scriptscriptstyle{#2} \cdot \scriptscriptstyle{#3}}}}
}%

\newcommandx*{\ce}[5][1=,2=,3=,4=,5=,usedefault]{%
  c^{\physe}_{
  \ifthenelse{\isempty{#1}}
  {}
  {
  \ifthenelse{\isempty{#2}}
  {#1}
  {\ifthenelse{\isempty{#4}}
  {
  #1 \cdot \scriptscriptstyle{#2} \cdot \scriptscriptstyle{#3}
  }
  {
  #1 \cdot \scriptscriptstyle{#2} \cdot \scriptscriptstyle{#3}\cdot \scriptscriptstyle{#4} \cdot \scriptscriptstyle{#5}}
  }
  }
  }
}%

\newcommandx*{\cf}[5][1=,2=,3=,4=,5=,usedefault]{%
  c^{\physf_{\varphi}}_{
  \ifthenelse{\isempty{#1}}
  {}
  {
  \ifthenelse{\isempty{#2}}
  {#1}
  {\ifthenelse{\isempty{#4}}
  {
  #1 \cdot \scriptscriptstyle{#2} \cdot \scriptscriptstyle{#3}
  }
  {
  #1 \cdot \scriptscriptstyle{#2} \cdot \scriptscriptstyle{#3}\cdot \scriptscriptstyle{#4} \cdot \scriptscriptstyle{#5}}
  }
  }
  }
}

\newcommandx*{\Cg}[3][1=,2=,3=,usedefault]{%
  C^{\physg}_{\ifthenelse{\isempty{#1}}{}{\ifthenelse{\isempty{#2}}{#1}{#1 \cdot \scriptscriptstyle{#2} \cdot \scriptscriptstyle{#3}}}}
}%

\newcommandx*{\Ce}[3][1=,2=,3=,usedefault]{%
  C^{\physe}_{\ifthenelse{\isempty{#1}}{}{\ifthenelse{\isempty{#2}}{#1}{#1 \cdot \scriptscriptstyle{#2} \cdot \scriptscriptstyle{#3}}}}
}%

\newcommandx*{\Cf}[5][1=,2=,3=,4=,5=,usedefault]{%
  C^{\physf_{\varphi}}_{
  \ifthenelse{\isempty{#1}}
  {}
  {
  \ifthenelse{\isempty{#2}}
  {#1}
  {\ifthenelse{\isempty{#4}}
  {
  #1 \cdot \scriptscriptstyle{#2} \cdot \scriptscriptstyle{#3}
  }
  {
  #1 \cdot \scriptscriptstyle{#2} \cdot \scriptscriptstyle{#3}\cdot \scriptscriptstyle{#4} \cdot \scriptscriptstyle{#5}}
  }
  }
  }
}

\newcommandx*{\Cx}[3][1=,2=,3=,usedefault]{%
  C^{\physx}_{\ifthenelse{\isempty{#1}}{}{\ifthenelse{\isempty{#2}}{#1}{#1 \cdot \scriptscriptstyle{#2} \cdot \scriptscriptstyle{#3}}}}
}%
\newcommandx*{\sx}[3][1=,2=,3=,usedefault]{%
  s^{\physx}_{\ifthenelse{\isempty{#1}}{}{\ifthenelse{\isempty{#2}}{#1}{#1 \cdot \scriptscriptstyle{#2} \cdot \scriptscriptstyle{#3}}}}
}%
\newcommandx*{\cx}[3][1=,2=,3=,usedefault]{%
  c^{\physx}_{\ifthenelse{\isempty{#1}}{}{\ifthenelse{\isempty{#2}}{#1}{#1 \cdot \scriptscriptstyle{#2} \cdot \scriptscriptstyle{#3}}}}
}%

\newcommandx*{\BetaC}[3][1=,2=,3=,usedefault]{%
  \beta\ag{#1}_{\ifthenelse{\isempty{#2}}{}{\cdot \scriptscriptstyle{#2} \cdot \scriptscriptstyle{#3}}}
}%

\newcommand{\physE}{{\mathpzc{E}}}

\newcommand{\vecset}[1]{\pr{\phys{#1}_i\ag{X}}_{i\in (1,2,3)}}
\newcommand{\vecsetf}{\pr{\phys{f}_{\varphi;i}\ag{X}}_{i\in (1,2,3)}}

\renewcommand{\u}[1]{\boldsymbol{#1}}

\newcommand{\p}{\,\textsf{\small pow}}

\newcommand{\m}[2]{
\IfEqCase{#2}{
       {i}{m_{#1\cdot #2}}%
   }[m_{#1\cdot {\scriptscriptstyle #2}}]}

\newcommand{\msub}[2]{
\IfEqCase{#2}{
      {i}{\, m_{#1\cdot #2}\,}%
  }[\,m_{#1\cdot {\scriptscriptstyle #2}}\,]}

\newcommand{\msubp}[2]{
\IfEqCase{#2}{
      {i}{\, \plus{m}_{#1\cdot #2}\,}%
  }[\,\plus{m}_{#1\cdot {\scriptscriptstyle #2}}\,]}

  \newcommand{\msubpp}[2]{
\IfEqCase{#2}{
      {i}{\, \plusplus{m}_{#1\cdot #2}\,}%
  }[\,\plusplus{m}_{#1\cdot {\scriptscriptstyle #2}}\,]}

\newcommand{\msubm}[2]{
\IfEqCase{#2}{
      {i}{\, \minus{m}_{#1\cdot #2}\,}%
  }[\,\minus{m}_{#1\cdot {\scriptscriptstyle #2}}\,]}

\newcommand{\msubmm}[2]{
\IfEqCase{#2}{
      {i}{\, \minusminus{m}_{#1\cdot #2}\,}%
  }[\,\minusminus{m}_{#1\cdot {\scriptscriptstyle #2}}\,]}

\newcommand{\gsub}[2]{g_{#1\cdot #2}}

\newcommand{\Wsub}[2]{
\IfEqCase{#2}{
      {i}{\, W_{#1\cdot #2}\,}%
  }[\,W_{#1\cdot {\scriptscriptstyle #2}}\,]}

\newcommand{\musub}[2]{\mu_{#1\cdot #2}}

\newcommand{\betasub}[2]{
\IfEqCase{#2}{
       {xxxx}{\,\beta_{ {#1}\cdot { #2}}\,}%
   }[\,\beta_{ {#1}\cdot {\scriptscriptstyle #2}}\,]}

   \newcommand{\Csub}[2]{
   \IfEqCase{#2}{
          {xxxx}{\,\beta_{ {#1}\cdot { #2}}\,}%
      }[\,S^{\pr{\physe}}_{ {#1}\cdot {\scriptscriptstyle #2}}\,]}

\newcommand{\alphasub}[2]{
\IfEqCase{#2}{
        {i}{\,\alpha_{ {#1}\cdot { \scriptstyle #2}}\,}%
    }[\,\alpha_{ {#1}\cdot {\scriptscriptstyle #2}}\,]}

\newcommand{\alphaBARsub}[2]{
\IfEqCase{#2}{
        {i}{\,\bar{\alpha}_{ {#1}\cdot { \scriptstyle #2}}\,}%
    }[\,\bar{\alpha}_{ {#1}\cdot {\scriptscriptstyle #2}}\,]}

\newcommand{\gammasub}[2]{
\IfEqCase{#2}{
        {i}{\,\gamma_{ {#1}\cdot { \scriptstyle #2}}\,}%
    }[\,\gamma_{ {#1}\cdot {\scriptscriptstyle #2}}\,]}

\newcommand{\Ksub}[2]{\,K_{ {#1}\cdot {\scriptscriptstyle #2}}\,}

\newcommand{\incylinder}{\in(1,2,3)}

\newcommand{\infour}{\in(1,\ldots,4)}

\newcommand{\Subs}[3]{
\IfEqCase{#3}{
        {i}{\,#1_{ {#2}\cdot { \scriptstyle #3}}\,}%
    }[\,#1_{ {#2}\cdot {\scriptscriptstyle #3}}\,]}

\usepackage{wrapfig}

\newcommand{\RN}[1]{%
  \textup{\uppercase\expandafter{\romannumeral#1}}%
}

  \DeclareMathAlphabet{\mathvz}{OT1}{LinuxBiolinumT-OsF}{m}{sl}

  \DeclareMathAlphabet{\mathlib}{OT1}{LinuxLibertineT-OsF}{m}{it}
  \DeclareMathAlphabet{\mathbio}{OT1}{LinuxBiolinumT-OsF}{m}{it}
\DeclareMathAlphabet{\mathpzc}{OT1}{pzc}{m}{it}

\newcommand*\rel@kern[1]{\kern#1\dimexpr\macc@kerna}
\newcommand*\widebar[1]{%
  \begingroup
  \def\mathaccent##1##2{%
    \rel@kern{0.8}%
    \overline{\rel@kern{-0.8}\macc@nucleus\rel@kern{0.2}}%
    \rel@kern{-0.2}%
  }%
  \macc@depth\@ne
  \let\math@bgroup\@empty \let\math@egroup\macc@set@skewchar
  \mathsurround\z@ \frozen@everymath{\mathgroup\macc@group\relax}%
  \macc@set@skewchar\relax
  \let\mathaccentV\macc@nested@a
  \macc@nested@a\relax111{#1}%
  \endgroup
}
\makeatother

\usepackage[prefix=sol]{xcolor-solarized}
\usepackage{booktabs}
\usepackage{pdflscape}
\usepackage{threeparttable}
\usepackage{tabto}

\usepackage{xspace}

\newcommand{\subf}[1]{\pr{\textsf{#1}}}

\usepackage{tablefootnote}

\usepackage{algorithm,algorithmicx}
\usepackage{algpseudocode}
\usepackage[hang,flushmargin]{footmisc}

\usepackage{xspace}

\usepackage{bm}

\usepackage{tabularx} 

\newcommand{\veryshortarrow}[1][3pt]{\mathrel{%
   \hbox{\rule[\dimexpr\fontdimen22\textfont2-.2pt\relax]{#1}{.4pt}}%
   \mkern-4mu\hbox{\usefont{U}{lasy}{m}{n}\symbol{41}}}}

\usepackage[font={tiny,it}]{caption}

\newcommand*{\doubleequation}[3][]{%
    \par\vskip\abovedisplayskip\noindent
    \if\relax\detokenize{#1}\relax
       \let\@dblLabelI\@empty
       \let\@dblLabelII\@empty
    \else 
       \@dblequationAux #1,%
    \fi
    \makebox[0.5\linewidth-1.5em]{%
     \hspace{\stretch2}%
     \makebox[0pt]{$\displaystyle #2$}%
     \hspace{\stretch1}%
    }%
    \makebox[0.5\linewidth-1.5em]{%
     \hspace{\stretch1}%
     \makebox[0pt]{$\displaystyle #3$}%
     \hspace{\stretch2}%
    }%
    \makebox[3em][r]{(%
  \refstepcounter{equation}\theequation\@dblLabelI,
  \refstepcounter{equation}\theequation\@dblLabelII)}%
  \par\vskip\belowdisplayskip
}
\makeatother

\begin{document}


\title{
Functional significance of lamellar architecture in marine sponge fibers: conditions for when  splitting a cylindrical tube into an assembly of tubes will decrease its bending stiffness}

\author{Sayaka Kochiyama, Benjamin Grossman-Ponemon, \\ and Haneesh Kesari}

\maketitle

\begin{abstract}
Numerous ingenious engineering designs and devices have been the product of bio-inspiration.  Bone, nacre, and other such stiff structural biological materials (SSBMs) are composites that contain mineral and organic materials interlaid together in layers. In nacre, this lamellar architecture is known to contribute to its fracture toughness, and has been investigated with the goal of discovering new engineering material design principles that can aid the development of synthetic composites that are both strong and tough. The skeletal anchor fibers (spicules) of \textit{Euplectella aspergillum} (\textit{Ea.}) also display a lamellar architecture; however, it was recently shown using fracture mechanics experiments and computations that the lamellar structure in them does not significantly contribute
 to their fracture toughness. An alternate hypothesis---load carrying capacity (LCC) hypothesis---regarding the lamellar architecture’s functional significance in \textit{Ea.} spicules is that it enhances the spicule’s strength, rather than its toughness. From the ecology perspective the LCC hypothesis is certainly plausible, since a higher strength would allow the spicules to more firmly anchor \textit{Ea.} to the sea floor, which would be beneficial to it since it is a filter feeding animal. In this paper we present support for the LCC hypothesis from the solid and structural mechanics perspective, which, compared to the support from ecology, is far harder to identify but equally, if not more, compelling and valid. We found that when the spicule functions in a knotted configuration a reduced bending stiffness benefits its load carrying capacity, and that for a wide class of materials that are consistent with the spicules' axial symmetry, sectioning a cylindrical tube into an assembly of co-axial tubes can reduce the tube’s bending stiffness. The mechanics theory developed in this paper has applications beyond providing support for the LCC hypothesis. For example, it makes apparent many design strategies for reducing the bending stiffness of large industrial cables (e.g., undersea optical data transmission cables)  while maintaining their tensile strength, which has benefits towards the handling, storage, and transportation of such cables.\\

\noindent\textit{Keywords}\\
\noindent structure property connection; curvilinear anisotropic elasticity; composites architectured materials; beams; material design

\end{abstract}

\section{Introduction}

\begin{figure}[t]
    \centering
    \graphicspath{{figure/}}
        \includegraphics[width=\textwidth]{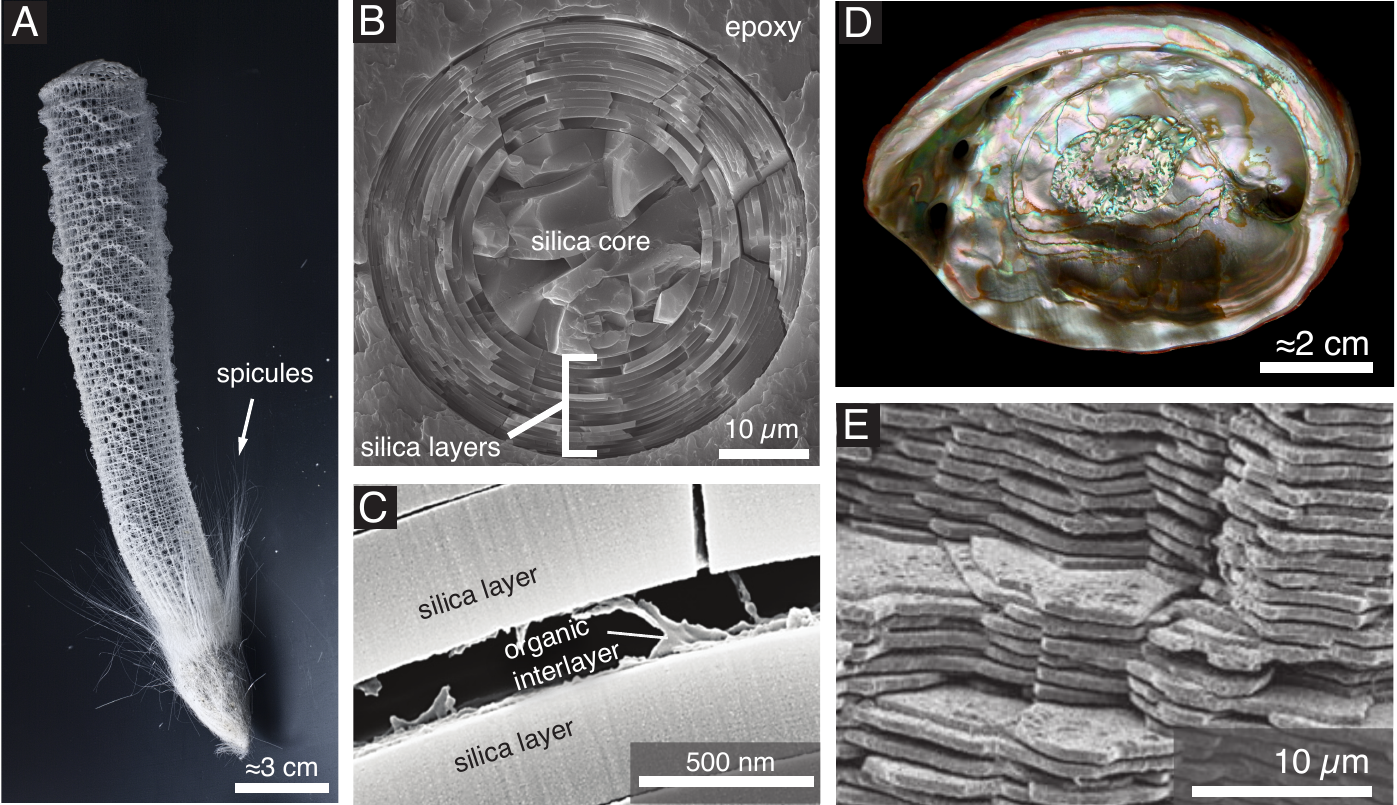}
    \caption{Stiff biological materials with lamellar architectures. 
$\subf{A}$ The entire skeletal structure of an \textit{Ea.} sponge (modiﬁed from ~\cite{monn2015new}; copyright 2015 National Academy of Sciences). 
The basalia spicules, which are identiﬁed with a white arrow, are around $50$ micrometers in diameter and can be several centimeters long. 
$\subf{B}$ A scanning electron microscope (SEM) image showing the cross section of an \textit{Ea.} basalia spicule reveals its cylindrically layered internal architecture (modiﬁed from~\cite{monn2015new}; copyright 2015 National Academy of Sciences). 
$\subf{C}$ An SEM image showing the organic layer that separates the adjacent silica layers (modified with permission from~\cite{weaver2007hierarchical}; copyright 2007 Elsevier).
$\subf{D}$ The shell of \textit{Haliotis rufescens}—the red abalone (image courtesy of John Varner). 
$\subf{E}$ An SEM image of nacre from \textit{H. rufescens} showing its brick-and-mortar layered architecture, where aragonite tablets correspond to the bricks and protein layers correspond to the mortar (modiﬁed with permission from~\cite{rabiei2012nacre}; copyright 2012 Royal Society of Chemistry).
  }
    \label{fig:schematic}
\end{figure}

Materials found in nature have been studied with the goal of deriving inspiration for the development of novel engineering structural materials. 
Spider silk, bone, antlers, tortoise carapaces, and fish scales are but a few among the list of structures and materials of biological origin that have been the subject of such bio-inspired material studies~\cite{chen2013bio,chen2012biological}. 
A recent addition to this list includes the skeleton and skeletal structures of the marine sponge \textit{Euplectella aspergillum} (\textit{Ea.}). 
Notably, its fiber-like structures that are composed primarily of silica, called the basalia spicules,  have attracted attention owing to its internal lamellar architecture (see Fig.\ref{fig:schematic}$\subf{A}, \subf{B}$). 
Basalia spicules act as anchors to hold the sponge onto the sea floor, and are roughly 50 micrometers in diameter and several centimeters in length~\cite{monn2015new}. 

Each fiber is an assembly of $\approx$25 co-axial tubular silica layers around a central, cylindrical core, where each adjacent pair of silica layers is separated by a thin, compliant organic interlayer (see Fig.\ref{fig:schematic}$\subf{C}$)~\cite{monn2015new, weaver2007hierarchical}. 
Lamellar architecture consisting of alternating stiff and compliant layers is also found in  nacre. In nacre the lamellar architecture is known to  enhance its fracture toughness~\cite{currey1977mechanical}. Therefore, it has been often assumed that the role of the \textit{Ea.} basalia spicules' internal lamellar architecture is to provide the basalia spicules with enhanced fracture toughness. 

However, recent direct fracture toughness measurements of the \textit{Ea.} basalia spicules showed that the basalia spicules' lamellar architecture does not provide them with any significant fracture toughness enhancement~\cite{monn2020lamellar}. 
This finding now reopens the question of what, if any, is the benefit of the basalia spicules' lamellar architecture to the sponge.

\begin{figure}[t]
    \centering
    \graphicspath{{figure/}}
        \includegraphics[width=0.8\textwidth]{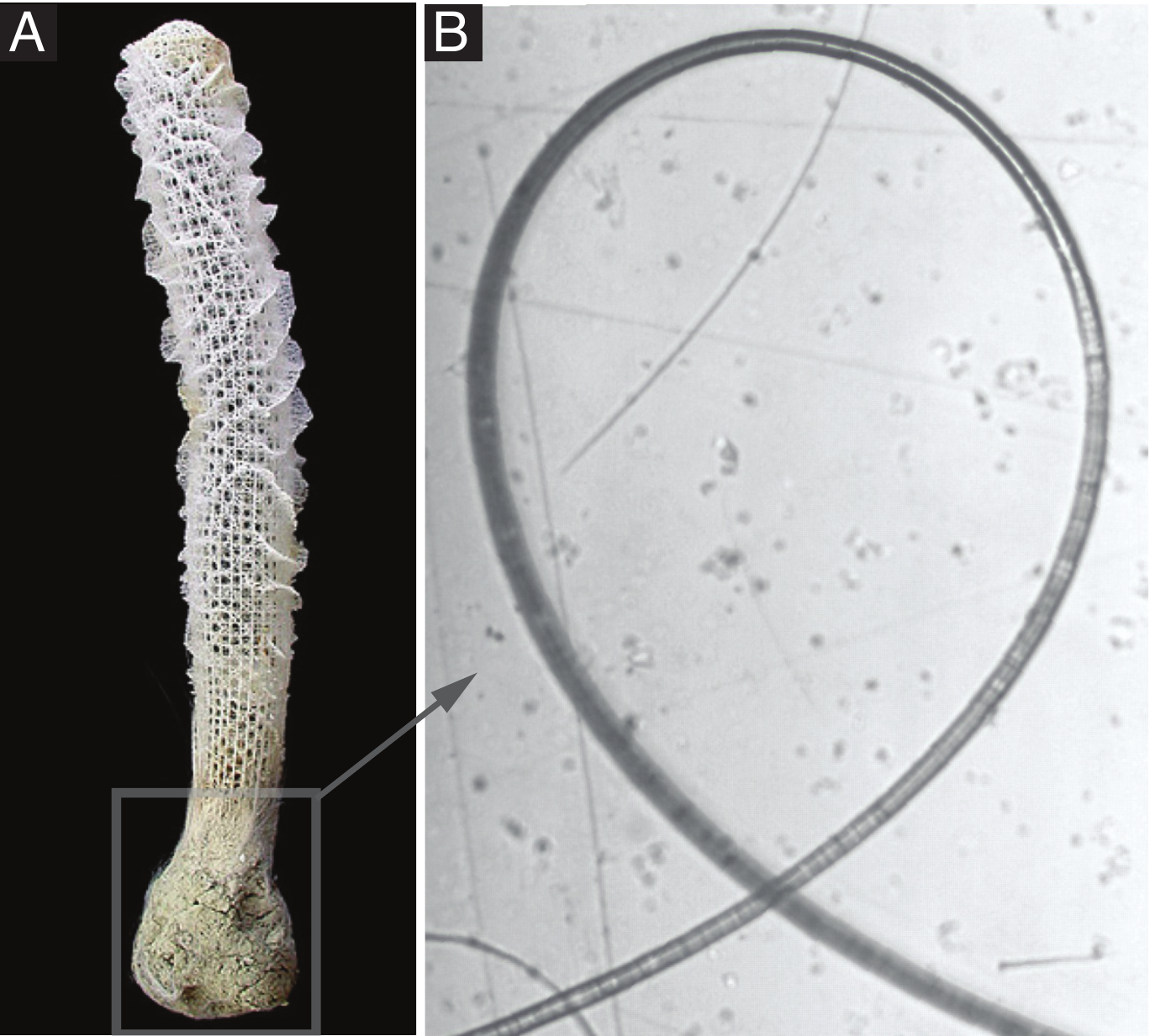}
    \caption{A \textit{Euplectella aspergillum} (\textit{Ea.}) basalia spicule in a looped configuration. (A) \textit{Ea.} basalia spicules anchor the sponge onto the sea floor. (B) \textit{Ea.} basalia spicules may form loops as they get tightened around sea floor sediments.
  }
    \label{fig:spiculeLoop}
\end{figure}

We refer to the maximum tensile force that a (\textit{Ea.} basalia) spicule is capable of transmitting along its length from the sea floor to the skeleton as its load carrying capacity (LCC).
The more tensile force that the spicules are capable of transmiting from the sea floor to the animal's skeleton the better they presumably serve their goal of anchoring the animal firmly to the ocean floor. 

This motivated us to put forward the hypothesis in a previous study~\cite{monn2015new} that the lamellar architecture's primary role is to contribute to the spicule’s LCC. 
From here on we will be referring to this hypothesis as the LCC hypothesis.
The primary evidence in support of the LCC hypothesis put forward in our previous paper~\cite{monn2015new} was the positive correlation between the experimentally measured  silica layer thicknesses and those in a spicule model in which the thicknesses were chosen to maximize the model's LCC.
In this paper, we present further arguments in support of the LCC hypothesis and develop a mechanics theory that can be used to more thoroughly check the validity of the LCC hypothesis once the spicule's elasticity becomes better characterized.

If a basalia spicule were to remain perfectly straight and have its terminal ends loaded uniformly across its cross-sections, then the only factor relevant to its LCC would be the strengths and thicknesses of its individual layers. 
It has been observed that the strength of ceramics increases with decreasing specimen thicknesses~\cite{griffith1921vi, mckinney1981specimen}.
Therefore, for the case of spicules that are subject to simple tensile loading we have immediate support for the LCC hypothesis.
However, by studying the arrangement of the spicules within the mud bulb that is often found at the base of \textit{Ea.} skeleton (see Fig.\ref{fig:spiculeLoop}$\subf{A}$), it is reasonable to conclude that more of the spicules function in a looped configuration within the sediment rather than in an unlooped configuration.
Therefore, the spicule's LCC in the looped configuration is as relevant as, if not more than, its capacity in an unlooped configuration.
The aim of this paper is to construct and present arguments that provide support for the LCC hypothesis  for the case in which the spicules function in a looped configuration (see Fig.\ref{fig:spiculeLoop}$\subf{B}$).

A nonlinear beam model for the spicule predicts that the looped configuration's LCC, in addition to the layers strengths and thicknesses, will also depend inversely on the  spicule's effective bending stiffness.\footnote{We will present this nonlinear beam model elsewhere.}

Atomic force microscopy testing \cite{weaver2007hierarchical} reveals that the  thin organic layers that separate the silicious layers are much more compliant compared to the silicious layers (see Fig.\ref{fig:schematic}$\subf{C}$). Motivated by this observation, in all our analysis as a simplifying assumption, we take that the organic layers do not transmit any shear stresses between the silica layers.

If we model the silica of the layers as being a homogenous, isotropic, linear elastic material, and model the overall deformation of the spicule using small deformation beam theories, then we find that the spicule's effective bending stiffness is independent of the number of silica layers. This result, when taken alone, weakens the possibility that (for the case of the looped configuration) the LCC hypothesis is true. Modeling the spicule silica using a homogenous isotropic material model seems natural considering that the spicule seems to have axial symmetry. That is, the spicule would look the same on rotating it by any angle about its long axis. However, there are many other material models, and not just the isotropic material model, that allow for the existence of axial symmetry. We introduce and discuss those other material models in  \S\ref{sec:MaterialSymmetries}. 

On considering those more general elastic material models, we found that the stiffness of a spicule can decrease with the introduction of layers.  These more general material models are called helically orthotropic material models (HOMM), and come under the general class of curvilinear anisotropic material models. Working under the paradigm of HOMM, we found the precise condition on the elastic constants of a cylindrical tube that on being satisfied guarantees that the cylindrical tube's bending stiffness will decrease on splitting the cylindrical tube into four or more co-axial tubes. Currently, the exact elastic characteristics of the material composing the spicules are not known. Therefore, our result that when the spicules' elasticity posseses certain curvilinear anisotropy characteristics then its bending stiffness will decrease with the number of layers provides support to the LCC hypothesis; this is because, as we mentioned previously, the looped configuration's LCC depends inversely on its bending stiffness.

The condition which on being satisfied guarantees that a cylindrical tube's bending stiffness will decrease on splitting the cylindrical tube into four or more co-axial tubes are given in \eqref{eq:nTC} in \S\ref{sec:NecessaryC}. In our numerical experiments we, in fact, found that  when  \eqref{eq:nTC} is satisfied the bending stiffness reduces even on splitting the tube into just two tubes.

We also derived the condition which on being satisfied guarantees that a cylindrical tube's bending stiffness will not decrease on being split into any number of layers. This condition is given in \eqref{eq:TC} in \S\ref{sec:TCsatisfied}. The isotropic material model always satisfies this condition. Thus the finding from the small deformation beam theories that the bending stiffness of an isotropic beam cannot be reduced by splitting it into two or more co-axial tubes is consistent with our result.

The outline of the paper is as follows.

In \S\ref{sec:mathprelim} we introduce the geometrical notions necessary for describing the various curvilinear anisotropic elasticity material models that we discuss in this paper. 
In \S\ref{sec:MechanicsPreliminaries} we briefly review the theory of curvilinear anisotropy and present the various 
curvilinear anisotropic elasticity material models.
We present three primary results in this paper in \S\ref{sec:Results}. 
In \S\ref{sec:BendingStiffness} we present a structural mechanics model for the spicules based on the work of Jolicouer and Cardour~\cite{Jolicoeur1994}. 
Each layer in our structural mechanics model for the spicule can be composed of a different helically orthotropic material.
In \S\ref{sec:NecessaryC} we present the condition on the elastic constants of a cylindrical tube that on being satisfied guarantees that the cylindrical tube’s bending stiffness
will decrease on splitting the tube into four or more co-axial tubes. 
In \S\ref{sec:TCsatisfied} we present the condition on the elastic constants of a cylindrical tube that on being satisfied guarantees that the cylindrical tube’s bending stiffness will not decrease on splitting the tube into any number of layers. 
In \S\ref{sec:Discussion} we discuss the numerical calculations that we carried out for checking the validity of the results we presented in \S\ref{sec:NecessaryC} and \S\ref{sec:TCsatisfied}.
Our arguments and results are based on several assumptions ranging from the mechanical behavior of the spicules to the role of shear stresses in bending stiffness. We collect and list the most pertinent assumptions in our work in  \S\ref{sec:conclusion}, so that it is easier to ascertain the validity of our arguments and results in the future.

\label{sec:intro}

\label{sec:bending_model}

\section{Mathematical preliminaries}
\label{sec:mathprelim}
\subsection{Geometrical notions} 
\label{subsec:CoordinateSystem}

We take that the spicule assumes its physical bent configurations in the space $\mathcal{E}_{\rm{R}}$, which is an  affine point space. 
We take the Euclidean space $\mathbb{E}_{\rm{R}}$ to be  $\mathcal{E}_{\rm{R}}$'s associated vector translation space. 
We refer to $\mathcal{E}_{\rm{R}}$ as the reference point space. 
We take some arbitrary point $O\in \mathcal{E}_{\rm{R}}$ to be $\mathcal{E}_{\rm{R}}$'s origin. 

\subsubsection{Basis vectors}
\label{subsec:basisvecs}
\paragraph{Cartesian basis vectors}
Let $\physx=\left(\physx_i\right)_{i\in (1,2,3)}$ be an arbitrary, orthonormal set of vectors in $\mathbb{E}_{\rm{R}}$. 
By orthornormal we mean that $\physx_i\cdot\physx_j = \delta_{ij}$, for~$i$,~$j\incylinder$, where $\delta_{ij}$ is the Kronecker delta symbol and is defined to be unity if~$i=j$ and naught otherwise. 
The Cartesian co-ordinates of $X\in \mathcal{E}_{\rm{R}}$, which we denote as $\breve{\mathsf{X}}\ag{X}=\pr{\breve{\mathsf{X}}_i\ag{X}}_{i\in (1,2,3)}$, are components of the  vector $X-O\in \mathbb{E}_{\rm{R}}$ with respect to $\physx_i$. 
We call the map $\mathcal{E}_{\rm{R}}$ $ \ni X \mapsto \breve{\mathsf{X}}\ag{X}\in \mathbb{R}^3$ the Cartesian co-ordinate map. 
There also exists the inverse Cartesian co-ordinate map $\mathbb{R}^3\ni \mathsf{X} \mapsto \breve{X}\ag{\mathsf{X}}\in \mathcal{E}_{\rm{R}}$, such that $\breve{X}\ag{\breve{\mathsf{X}}\ag{X}}=X$.

\paragraph{Cylindrical basis vectors} The cylindrical co-ordinates of $X$, which we denote as $(\breve{r}\ag{X}, \breve{\theta}\ag{X}, \breve{z}\ag{X})$, are defined in the standard manner using the Cartesian co-ordinates $\breve{\mathsf{X}}\ag{X}$. 
Let $\physe\ag{X}:=\pr{\physe_{i}\ag{X}}_{i\in (1,2,3)}$, where
\begin{subequations}
\begin{align}
\physe_{1}\ag{X} & = \cos\ag{\breve{\theta}\ag{X}} \physx_1 + \sin\ag{\breve{\theta}\ag{X}} \physx_2 , \\
\physe_{2}\ag{X} & = -\sin\ag{\breve{\theta}\ag{X}} \physx_1 + \cos\ag{\breve{\theta}\ag{X}} \physx_2 , \\
\physe_{3} \ag{X} & =  \physx_3.
\end{align}
\end{subequations} 
The vectors $\physe\ag{X}$ are simply the cylindrical  basis vectors.

\paragraph{Helical basis vectors}
Let $\physf_{\varphi}\ag{X}:=\pr{\physf_{\varphi;i}\ag{X}}_{i\in (1,2,3)}$, where
\begin{equation}
\physf_{\varphi;i} \ag{X} = \sum_{j \incylinder } Q_{\dt i\dt j}\ag{\varphi}\physe_{j}\ag{X},
\label{eq:Qtransform}
\end{equation}
\begin{equation}
Q_{\dt \dt}\ag{\varphi}
=
\left(\begin{array}{ccc}
1 & 0 & 0 \\ [3 pt]
 0 & \cos\ag{\varphi} & -\sin\ag{\varphi} \\ [3 pt]
 0 & \sin\ag{\varphi} & \cos\ag{\varphi}
\end{array}
\right),
\label{eq:QDef}
\end{equation} 
$Q_{\dt \dt}\ag{\varphi} \in \mathcal{M}_{6\times 6}\pr{\mathbb{R}}
$\footnote{We consider a matrix to be an ordered set of elements where all the elements are ordered sets of the same cardinality. Specially, we denote the  ordered set containing $m\in \mathbb{Z}_{\ge 1}$ elements where each element is an ordered set containing $n\in\mathbb{Z}_{\ge 1}$ real numbers as $\mathcal{M}_{m\times n}\pr{\mathbb{R}}$.} and $\varphi \in [0,\pi]$. 
We refer to $\physf_{\varphi}\ag{X}$ as the helical basis vectors, and to $\varphi$ as the helix angle. The reason behind this is as follows.

Consider the space curve $\breve{\Gamma}_{X; \varphi}\ag{\cdot}: \mathbb{R}\to \mathbb{R}^3$,
$$
\breve{\Gamma}_{X; \varphi}\ag{\xi}=\pr{ \breve{r}\ag{X}\cos\ag{\breve{\theta}\ag{X}-\xi},\breve{r}\ag{X}\sin\ag{\breve{\theta}\ag{X}-\xi}, \breve{r}\ag{X}\xi \cot\ag{\varphi}+\breve{z}\ag{X}}.
$$ 
The curve $\breve{\Gamma}_{X; \varphi}\ag{\cdot}$ is a helix in $\mathbb{R}^3$ with radius $\breve{r}\ag{X}$ and pitch $2 \pi \breve{r}\ag{X}/\tan\ag{\varphi}$ that passes through $\breve{\mathsf{X}}\ag{X}$. 
We define the reference helix corresponding to $\breve{\Gamma}_{X; \varphi}\ag{\cdot}$ as $\breve{X}\circ\breve{\Gamma}_{X; \varphi}\ag{\cdot}: \mathbb{R}\to \mathcal{E}_{\rm R}$.
The reference helix lies in $\mathcal{E}_{\rm R}$ and passes through the point $X$. 
The vectors $\physf_{\varphi}\ag{X}$ are closely related to the Frenet-Serret frame~\cite{forsyth1912lectures} of the reference helix at $X$. 
More specifically, the tangent vector in the Frenet-Serret frame of the reference helix at $X$ equals $\physf_{\varphi;3}\ag{X}$, the normal vector equals $-\physf_{\varphi;1}\ag{X}$, and the bi-normal vector equals $-\physf_{\varphi;2}\ag{X}$.

The vector sets $\vecset{e}$ and $\vecsetf$ are, respectively, orthonormal. 
The set $\vecsetf$ can be obtained by rotating $\vecset{e}$ about $\physe_1\ag{X}$ (the radial vector) by $-\varphi$ (i.e., $\varphi$ in the clockwise direction).

\begin{figure}[t]
    \centering
    \graphicspath{{figure/}}

        \includegraphics[width=\textwidth]{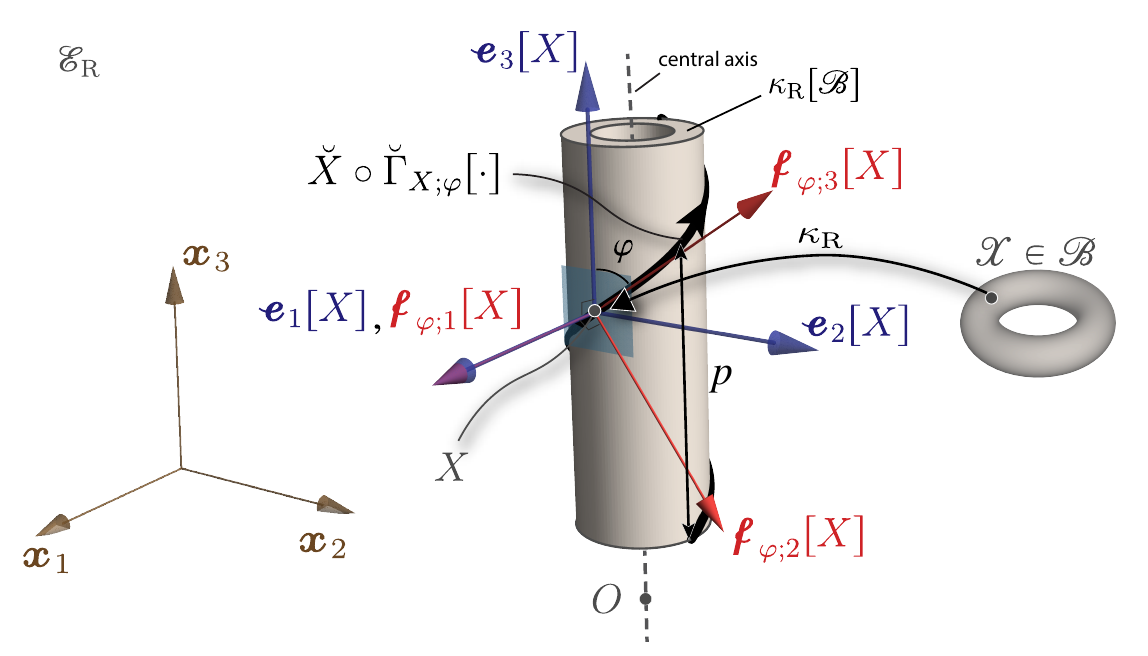}
    \caption{
  The geometry of a single cylindrical layer in our model for the spicule.
  Our spicule model consists of one or more of such cylindrical layers. 
  The various geometrical quantities illustrated and marked, such as $\mathcal{B}$, $\physe_i\ag{X}$, $i=1,2,3$, $\varphi$, $\breve{X}\circ\breve{\Gamma}_{X;\varphi}\ag{\cdot}$, etc., are all defined and discussed in \S\ref{subsec:basisvecs} and \S\ref{sec:LinearElasticity}.  
  }

    \label{fig:cylinderSchematic}
\end{figure}

\subsubsection{Units}
\label{sec:units}

Following the formalism introduced in~\cite{rahaman2020accelerometer,deng2021angle}, we take the vectors in $\mathbb{E}_{\rm R}$ to have units of length, such as meters or millimeters, and refer to $\mathbb{E}_{\rm R}$ as the physical matter vector space. 
Consequently, the vectors in the sets $\pr{\physx_i}_{i\in (1,2,3)}$, $\vecset{e}$, and $\vecsetf$ all carry units of length, and the components of a vector in $\mathbb{E}_{\rm R}$ with respect to any of these sets are dimensionless.  

In fact, as per our formalism, in all our physical (Euclidean) vector spaces units will be an intrinsic aspect of the spaces' vectors themselves. 
And all elements of a physical vector space have the same units. 
For example, say that $\mathbb{F}$ is the force Euclidean vector space, and $\physf_1$, $\physf_2$, $\physf_3$ are an orthonormal set of vectors in $\mathbb{F}$. 
Then,  $\physf_i$ carry with them units of force. 
These units can be \textsf{Newton},  \textsf{milli-Newton}, etc. 
Consequenctly, if an arbitrary vector $\physf\in \mathbb{F}$ is equal to $\sum_{i\in(1,2,3)}f_i\physf_i$, then the components $f_i$ are dimensionless, i.e., they belong to $\mathbb{R}$. 

We use the map $U\ag{\cdot}$ to explicitly refer to a vector space's units.
That is, it takes a physical vector space as an argument and  returns the units carried by the elements of that space. 
For example, $U\ag{\mathbb{E}_{\rm R}}$ can be $\textsf{meters}$, $\textsf{micro-meters}$, etc., and  $U\ag{\mathbb{F}}$ can be \textsf{Newtons}, \textsf{mill-newtons}, etc.

\section{Mechanics preliminaries\label{sec:MechanicsPreliminaries}}

Let $\mathcal{B}$ be a manifold homeomorphic to the topological space formed by sweeping a disk around a circle (solid torus). 
We call $\mathcal{B}$ the material manifold and call its points material particles. 
Let $\kappa_{R}:\mathcal{B}\to \mathcal{E}_{\rm R}$ be a continuous injective map such that $\kappa_{R}\ag{\mathcal{B}}$ is a
straight tube having an annular cross-section with  its axis passing through $O$ and parallel to the $\physx_3$ direction (see Fig.\ref{fig:cylinderSchematic}). 
We call $\kappa_{R}$ the reference configuration, and $\kappa_{R}\ag{\mathcal{B}}$ the reference body.

\subsection{Linear elasticity\label{sec:LinearElasticity}}

As per the generalized Hooke's law
\begin{equation}
\u{\sigma}\ag{\mathcal{x}} =  \mathbbm{c}\ag{\mathcal{x}}\, \u{\epsilon}\ag{\mathcal{x}},
\label{eq:GeneralizedHooke}
\end{equation}
where $\mathcal{x}\in \mathcal{B}$ is an arbitrary material particle and $\u{\sigma}\ag{\mathcal{x}}$,~$\u{\epsilon}\ag{\mathcal{x}}$, and~$\mathbbm{c}\ag{\mathcal{x}}$ are, respectively, the Cauchy stress tensor, the infinitesimal strain tensor, and the elastic stiffness tensor at $\mathcal{x}$.
The Hooke's law can also be expressed as
\begin{equation}
 \u{\epsilon}\ag{\mathcal{x}} = \mathbbm{s}\ag{\mathcal{x}}\, \u{\sigma}\ag{\mathcal{x}},
\label{eq:GeneralizedHookeInverse}
\end{equation}
where~$\mathbbm{s}\ag{\mathcal{x}}$ is the elastic compliance tensor, which is the inverse of $\mathbbm{c}\ag{\mathcal{x}}$.

\paragraph{Compliance tensor components}
We take $\physg\ag{\mathcal{x}}:=:\pr{\physg_i\ag{\mathcal{x}}}_{i\in (1,2,3)}$ to be an arbitrary set of orthonormal vectors in $\mathbb{E}_{\rm R}$. 
For example, it can be $\physe\ag{\mathcal{x}}$ or  $\physf_{\varphi}\ag{\mathcal{x}}$. 
We denote the components of $\mathbbm{s}\ag{\mathcal{x}}$ with respect to bases related to $\physg\ag{\mathcal{x}}$ as $\sgC{i}{j}{k}{l}\ag{\mathcal{x}}$, where $i,j,k,l \in (1,2,3)$. 
In general $\sgC{i}{j}{k}{l}\ag{\cdot}$ denote 81 real valued functions over $\mathcal{B}$. 
It is a standard exercise to show that these 81 functions  are not independent and can be  expressed using only 21 functions. 
We denote those 21 functions as $\sgM{i}{j}\ag{\cdot}:\mathcal{B}\to \mathbb{R}$, $i\in (1,\ldots,6)$ and $1\le  j\le i $. 
We also define the functions  $\sgM{i}{j}\ag{\cdot}$ where $i\in (1,\ldots, 6)$ and $i<j\le 6$ as $\sgM{j}{i}\ag{\cdot}$. 
The numbers $\sgC{i}{j}{k}{l}\ag{\mathcal{x}}$  and $\sgM{i}{j}\ag{\mathcal{x}}$ are related to each other as
\begin{align}
\sgC{i}{j}{k}{l}\ag{\mathcal{x}}
&=
\sgM{{\mathsf{voi}}\ag{i,j}}{\mathsf{voi}\ag{k,l}}\ag{\mathcal{x}},
\label{eq:sCompVoigt}
\end{align}
where $\textsf{voi}:(1,2,3)^2\to (1,\ldots,6)$,
$\textsf{voi}\ag{i,i}=i$, $\textsf{voi}\ag{2,3}=4$, $\textsf{voi}\ag{1,3}=5$, $\textsf{voi}\ag{1,2}=6$, and when $j<i$, $\textsf{voi}\ag{i,j}=\textsf{voi}\ag{j,i}$. 
As the reader might have deduced the function $\textsf{voi}\ag{\cdot}$ implements the Voigt notation.

\paragraph{Compliance matrices}
We term the symmetric matrix  $\sg\ag{\mathcal{x}} :=\pr{\sgM{i}{j}\ag{\mathcal{x}}}_{i,j\in (1,\ldots, 6)}$ the compliance matrix. 
In particular, we refer to $\sx\ag{\mathcal{x}}$, $\se\ag{\mathcal{x}}$, and
$\sf\ag{\mathcal{x}}$, respectively, as the Cartesian compliance matrix, the cylindrical compliance matrix, and the helical compliance matrix at $\mathcal{x}$.

\paragraph{Stiffness tensor components} 
We denote the components of $\mathbbm{c}\ag{\mathcal{x}}$ with respect to bases related to $\physg\ag{\mathcal{x}}$ as $\cgC{i}{j}{k}{l}\ag{\mathcal{x}}$, where $i,j,k,l \in (1,2,3)$. 
Similar to the case of compliance tensor components, the functions $\cgC{i}{j}{k}{l}\ag{\cdot}:\mathcal{B}\to \mathbb{R}$ can be  expressed using only 21 functions, which we denote as $\cgM{i}{j}\ag{\cdot}:\mathcal{B}\to \mathbb{R}$, $i\in (1,\ldots,6)$ and $1\le j\le i$. 
We also define the functions  $\cgM{i}{j}\ag{\cdot}$ where $i\in (1,\ldots, 6)$ and $i<j\le 6$ as $\cgM{j}{i}\ag{\cdot}$. 
The numbers $\cgC{i}{j}{k}{l}\ag{\mathcal{x}}$ and $\cgM{i}{j}\ag{\mathcal{x}}$ are related to each other as
\begin{align}
\cgC{i}{j}{k}{l}\ag{\mathcal{x}}
&=
\cgM{{\mathsf{voi}}\ag{i,j}}{\mathsf{voi}\ag{k,l}}\ag{\mathcal{x}}.
\label{eq:cCompVoigt}
\end{align}

\paragraph{Stiffness matrices} 
We term the symmetric matrix  $\cg \ag{\mathcal{x}}:=\pr{\cgM{i}{j}\ag{\mathcal{x}}}_{i,j\in (1,\ldots, 6)}$ the stiffness matrix. 
In particular, we refer to $\cx\ag{\mathcal{x}}$, $\ce \ag{\mathcal{x}}$, and
$\cf \ag{\mathcal{x}}$  as, respectively, the Cartesian stiffness matrix, the cylindrical stiffness matrix, and the helical stiffness matrix at $\mathcal{x}$.

\paragraph{Inverse stiffness matrices ($\Cg \ag{\mathcal{x}}$)\protect\footnote{In many cases, what we refer to as inverse stiffness matrices are referred to as compliance matrices; one instance of this is the work by Jolicoeur and Cardou~\cite{Jolicoeur1994}.}}
We define the inverse stiffness matrix $\Cg \ag{\mathcal{x}}=\textsf{Inv}\ag{\cg \ag{\mathcal{x}}}$. 
Here, $\textsf{Inv}\ag{\cdot}$ is the standard matrix inversion operation. 
Thus, $\Cg \ag{\mathcal{x}}$ belongs to the set of $6\times 6$ matrices of real numbers, $\mathcal{M}_{6\times 6}\pr{\mathbb{R}}$. 
In particular,  $\Cx \ag{\mathcal{x}}$, $\Ce \ag{\mathcal{x}}$, and $\Cf \ag{\mathcal{x}}$ are, respectively, by definition $\textsf{Inv}\ag{\cx \ag{\mathcal{x}}}$, $\textsf{Inv}\ag{\ce \ag{\mathcal{x}}}$, and $\textsf{Inv}\ag{\cf \ag{\mathcal{x}}}$. 
We refer to $\Cx \ag{\mathcal{x}}$, $\Ce \ag{\mathcal{x}}$, and $\Cf \ag{\mathcal{x}}$, respectively, as the Cartesian inverse  stiffness matrix, the cylindrical inverse  stiffness matrix, and the helical inverse  stiffness matrix at $\mathcal{x}$.

The procedure for computing $\pr{\sgC{i}{j}{k}{l}\ag{\mathcal{x}}}_{i,j,k,l\in (1,2,3)}$, $\sg\ag{\mathcal{x}}$,  $\pr{\cgC{i}{j}{k}{l}\ag{\mathcal{x}}}_{i,j,k,l\in (1,2,3)}$, $\cg\ag{\mathcal{x}}$, and $\Cg\ag{\mathcal{x}}$ from one another is outlined in Fig.\ref{figure:link}.

\subsection{Material models\label{sec:MaterialSymmetries}}

If $\Cg\ag{\cdot}$ is a constant function then we  say that $\mathcal{B}$ has $\physg$-homogeneity. 
More specifically, if $\Ce\ag{\cdot}$ is a constant function then we say that $\mathcal{B}$ has cylindrical  homogeneity, and if $\Cf\ag{\cdot}$ is a constant function then we  say that $\mathcal{B}$ has helical homogeneity.
Most applications of linear elasticity restrict themselves to the case where $\Cx\ag{\cdot}$  is a constant. 
This is the case of $\mathcal{B}$ being "homogeneous."
We will refer to this case as $\mathcal{B}$ having Cartesian homogeneity. 

We note that when $\Cg\ag{\cdot}$ is a constant function, $\cg\ag{\cdot}$, $\cgC{i}{j}{k}{l}\ag{\cdot}$, $\sg\ag{\cdot}$, and $\sgC{i}{j}{k}{l}\ag{\cdot}$ are constant functions as well.

\subsubsection{Cylindrically, helically, and Cartesian orthotropic, transversely-isotropic, cubic, and isotropic materials\label{sec:CylindricallyHelicalSymmetries}} 

\paragraph{Cylindrically, helically, and Cartesian orthotropic materials} 

We say that $\mathcal{B}$ is orthotropic iff in addition to $\Cg\ag{\cdot}$  being a constant function, its constant value has the form
\begin{equation}
\left(\begin{array}{cccccc}
\frac{1}{ E_{1} } & - \frac{\nu_{12}}{E_{1}} & -\frac{\nu_{13}}{E_{1}} & 0 & 0 & 0 \\[5 pt]
- \frac{\nu_{12}}{E_{1}} &\frac{ 1 }{ E_{2} }& - \frac{ \nu_{23} }{ E_{2} } & 0 & 0 & 0 \\[5 pt]
-\frac{\nu_{13}}{E_{1}} & - \frac{ \nu_{23} }{ E_{2} }& \frac{1}{ E_{3} }& 0 & 0 & 0 \\[5 pt]
0 & 0 & 0 & \frac{1}{ \mu_{23} }& 0 & 0 \\[5 pt]
0 & 0 & 0 & 0 & \frac{1}{ \mu_{13} }& 0 \\[5 pt]
0 & 0 & 0 & 0 & 0 & \frac{1}{\mu_{12}}
\end{array}\right),
\label{eq:Ortho}
\end{equation}
where $E_{1}, E_{2}, E_{3}, \nu_{12}, \nu_{13}, \nu_{23}, \mu_{12}, \mu_{13}$, and $\mu_{23}$ are real constants such that the matrix \eqref{eq:Ortho} is positive definite.
We denote the set of material properties $(E_{1}, E_{2}, E_{3},\allowbreak \nu_{12}, \nu_{13}, \nu_{23}, \mu_{12}, \mu_{13}, \mu_{23})$ as $\acute{M}$. 
In particular, if $\physg\ag{\mathcal{x}}=\physe\ag{\mathcal{x}}$ then we say that $\mathcal{B}$ is cylindrically orthotropic ($\physe$-orthotropic), if $\physg\ag{\mathcal{x}}=\physf_{\varphi}\ag{\mathcal{x}}$ then that it is  helically orthotropic ($\physf_{\varphi}$-orthotropic), and if $\physg\ag{\mathcal{x}}=\physx$ then that it is Cartesian orthotropic ($\physx$-orthotropic).

\paragraph{Cylindrically, helically, and Cartesian transversely isotropic materials} 

We say that $\mathcal{B}$ is transversely isotropic iff in addition to $\Cg\ag{\cdot}$ being a constant function its constant value has the form
\begin{equation}
 \left(
\begin{array}{cccccc}
 \frac{1}{E_p} & -\frac{\nu _p}{E_p} & -\frac{\nu_{pt}}{E_p} & 0 & 0 & 0 \\[5 pt]
 -\frac{\nu _p}{E_p} & \frac{1}{E_p} & -\frac{\nu_{pt}}{E_p} & 0 & 0 & 0 \\[5 pt]
 -\frac{\nu_{pt}}{E_p} & -\frac{\nu _{pt}}{E_p} & \frac{1}{E_t} & 0 & 0 & 0 \\[5 pt]
 0 & 0 & 0 & \frac{1}{\mu _t} & 0 & 0 \\[5 pt]
 0 & 0 & 0 & 0 & \frac{1}{\mu _t} & 0 \\[5 pt]
 0 & 0 & 0 & 0 & 0 & \frac{2 \left(\nu _p+1\right)}{E_p}
\end{array}
\right),
\label{eq:TISO}
\end{equation}
where $E_p$, $E_t$, $\nu_{p}$, $\nu_{pt}$, and $\mu_t$ are real constants such that the matrix \eqref{eq:TISO} is positive definite. 
We denote the set of material properties $\pr{E_p, E_t, \nu_{p}, \nu_{pt}, \mu_t}$ as $\tilde{M}$.
In particular, if $\physg\ag{\mathcal{x}}=\physe\ag{\mathcal{x}}$ then we say that $\mathcal{B}$ is cylindrically transversely isotropic ($\physe$-transversely isotropic), if $\physg\ag{\mathcal{x}}=\physf_{\varphi}\ag{\mathcal{x}}$ then that it is helically transversely isotropic ($\physf_{\varphi}$-transversely isotropic), and if $\physg\ag{\mathcal{x}}$ =$\boldsymbol{\mathscr{x}}$ then that it is Cartesian transversely isotropic ($\physx$-transversely isotropic).

\paragraph{Cylindrically, helically, and Cartesian cubic materials} 
We say that $\mathcal{B}$ is cubic iff in addition to $\Cg\ag{\cdot}$ being a constant function its constant value has the form
\begin{equation}
\left(
      \begin{array}{cccccc}
      \frac{1}{E_c} &-\frac{\nu_c}{E_c}& -\frac{\nu_c}{E_c} & 0 & 0 & 0 \\[5 pt]
       -\frac{\nu_c}{E_c} &\frac{1}{E_c}& -\frac{\nu_c}{E_c} & 0 & 0 & 0 \\[5 pt]
       -\frac{\nu_c}{E_c} &-\frac{\nu_c}{E_c}&\frac{1}{E_c} & 0 & 0 & 0 \\[5 pt]
       0 & 0 & 0 & \frac{1}{\mu_c} & 0 & 0 \\[5 pt]
       0 & 0 & 0 & 0 & \frac{1}{\mu_c} & 0 \\[5 pt]
       0 & 0 & 0 & 0 & 0 &  \frac{1}{\mu_c}
      \end{array}
  \right),
\label{eq:cubic}
\end{equation}
where $E_c$, $\nu_c$, and $\mu_c$ are real constants such that the matrix \eqref{eq:cubic} is positive definite. 
We denote the set of material properties $\pr{E_c, \nu_c, \mu_c}$ as $\grave{M}$.
In particular, if $\physg\ag{\mathcal{x}}=\physe\ag{\mathcal{x}}$ then we say that $\mathcal{B}$ is cylindrically cubic ($\physe$-cubic), if $\physg\ag{\mathcal{x}}=\physf_{\varphi}\ag{\mathcal{x}}$ then that it is helically cubic ($\physf_{\varphi}$-cubic), and if $\physg\ag{\mathcal{x}}$ = $\boldsymbol{\mathscr{x}}$ then that it is Cartesian cubic ($\physx$-cubic).

\paragraph{Cylindrically, helically, and Cartesian isotropic materials}

We say that $\mathcal{B}$ is isotropic iff in addition to $\Cg\ag{\cdot}$ being a constant function, its constant value has the form
\begin{equation}
\left(
      \begin{array}{cccccc}
      \frac{1}{E} &-\frac{\nu}{E}& -\frac{\nu}{E} & 0 & 0 & 0 \\[5 pt]
       -\frac{\nu}{E} &\frac{1}{E}& -\frac{\nu}{E} & 0 & 0 & 0 \\[5 pt]
       -\frac{\nu}{E} &-\frac{\nu}{E}&\frac{1}{E} & 0 & 0 & 0 \\[5 pt]
       0 & 0 & 0 & \frac{2 \left(\nu+1\right)}{E}& 0 & 0 \\[5 pt]
       0 & 0 & 0 & 0 &  \frac{2 \left(\nu +1\right)}{E}& 0 \\[5 pt]
       0 & 0 & 0 & 0 & 0 &  \frac{2 \left(\nu +1\right)}{E}
      \end{array}
  \right),
  \label{eq:isomatrix}
\end{equation}
where $E$  and $\nu$ are real constants such that the matrix \eqref{eq:isomatrix} is positive definite. 
We denote the set of material properties $\pr{E,\nu}$ as $\bar{M}$.
In particular, if $\physg\ag{\mathcal{x}}=\physe\ag{\mathcal{x}}$ then we say that $\mathcal{B}$ is cylindrically isotropic ($\physe$-isotropic), if $\physg\ag{\mathcal{x}}=\physf_{\varphi}\ag{\mathcal{x}}$ then that it is helically isotropic ($\physf_{\varphi}$-isotropic), and if $\physg\ag{\mathcal{x}}$ = $\boldsymbol{\mathscr{x}}$ then that it is Cartesian isotropic ($\physx$-isotropic).

\subsubsection{Interdependence between the materials models}
\label{sec:mmInterdependence}

\usetikzlibrary{math}
\tikzmath{\x= 1.8; \y=2.2; \yp=2.5;}

\begin{figure}
    \begin{center}
    \resizebox{\textwidth}{!}{%
        \begin{tikzpicture}[auto,
                       > = Stealth,
           node distance = 5cm and 10cm,
              box/.style = {draw=gray, very thick,
                            minimum height=11mm, text width=22mm,
                            align=center},
       every edge/.style = {draw, <->, very thick},
every edge quotes/.style = {font=\footnotesize, align=center, inner sep=1pt},
decoration={markings,
    mark=at position .25 with
      with { \node [cross out,draw=red, inner sep=10pt, line width=1pt] {};}}
                            ]
    \node (He-Ortho) at (0, 0) [box]{$\physf$-Orthotropic};
    \node (He-TransIso) at (-\x, \yp) [box]{$\physf$-Transversely isotropic};
    \node (He-Cubic) at (\x, \yp) [box]{$\physf$-Cubic};
     \node (He-Iso) at (0, \yp + \y) [box] {$\physf$-Isotropic};
     \node (Cyl-Ortho) at (0, -\y )[box] {$\physe$-Orthotropic};
    \node (Cyl-TransIso) at (-3*\x, \yp) [box]{$\physe$-Transversely isotropic};
     \node (Cyl-Cubic) at (3*\x, \yp) [box] {$\physe$-Cubic};
    \node (Cyl-Iso) at (0, \yp+2*\y) [box] {$\physe$-Isotropic};
     \node (Cart-Iso) at (0, \yp+3*\y) [box]{$\physx$-Isotropic};
     \node (Cart-Ortho) at (0, -2*\y) [box] {$\physx$-Orthotropic};
     \node (Cart-TransIso) at (-5*\x, \yp) [box]{$\physx$-Transversely isotropic};
    \node (Cart-Cubic) at (5*\x, \yp) [box] {$\physx$-Cubic};

    \draw[->,  line width = 1 pt] (He-Iso.west)-|(He-TransIso.north);
    \draw[->,  line width = 1 pt] (He-Iso.east)-|(He-Cubic.north);
    \draw[->,line width = 1 pt] (He-TransIso.south)|-(He-Ortho.west);
    \draw[->, line width = 1 pt] (He-Cubic.south)|-(He-Ortho.east);

    \draw[->,  line width = 1pt, dashed] (Cyl-Iso.west) -| (Cyl-TransIso.north);
    \draw[->,  line width = 1pt, dashed] (Cyl-Iso.east)-|(Cyl-Cubic.north);
    \draw[->, line width = 1pt, dashed] (Cyl-TransIso.south)|-(Cyl-Ortho.west);
    \draw[->,  line width = 1pt, dashed] (Cyl-Cubic.south)|-(Cyl-Ortho.east);

    \draw[->, line width = 0.8 pt, dotted] (Cart-Iso.west)-|(Cart-TransIso.north);
    \draw[->,  line width = 0.8 pt, dotted] (Cart-Iso.east)-|(Cart-Cubic.north);
    \draw[->, line width = 0.8 pt, dotted] (Cart-TransIso.south)|-(Cart-Ortho.west);
    \draw[->,  line width = 0.8 pt, dotted] (Cart-Cubic.south)|-(Cart-Ortho.east);

    \draw[-{Kite[open,scale=1.2]}, line width = 1pt] (Cyl-TransIso)--(He-TransIso);
    \draw[-{Kite[open,scale=1.2]},line width = 1pt] (Cyl-Cubic)--(He-Cubic);
    \draw[-{Kite[open,scale=1.2]},line width=1pt] (Cyl-Ortho)--(He-Ortho);

    \draw[-{Triangle[open]},line width=1pt] (Cyl-Iso.70)--(Cart-Iso.290);
    \draw[-{Triangle[open]},line width=1pt] (Cart-Iso.250)--(Cyl-Iso.110);

    \draw[-{Kite[open,scale=1.2]},line width=1pt] (He-Iso.70)--(Cyl-Iso.290);
    \draw[-{Kite[open,scale=1.2]},line width=1pt] (Cyl-Iso.250)--(He-Iso.110);

    \draw[line width=2.5pt] ($(Cart-Iso.north west)+(-0.3,0.4)$)  rectangle ($(He-Iso.south east)+(0.3,-0.4)$);

    \draw[line width=2.5pt] ($(Cart-TransIso.north west)+(-0.4,0.3)$)  rectangle ($(Cyl-TransIso.south east)+(0.4,-0.3)$);

    \draw[-{Triangle[open]},line width=1pt] (Cart-TransIso.10)--(Cyl-TransIso.170);
    \draw[-{Triangle[open]},line width=1pt] (Cyl-TransIso.190)--(Cart-TransIso.350);

         \end{tikzpicture}
         }
         \end{center}

\caption{Interdependence between the material models considered in this paper. The following observations can be made: (A) It can be seen that within a set of material models that has $\physg$-homogeneity, the isotropic,  cubic, and transversely isotropic material models are all special cases of the orthotropic material model.  This is marked with solid arrows (\full)  for material models with helical homogeneity,  with  dashed arrows(\dashed) for those with cylindrical homogeneity, and with dotted arrows (\dotted) for those with Cartesian homogeneity. (B) Each arrow with a kite tip (\openkite)  marks the relationship between a helically homogeneous material model and a cylindrically homogeneous material. For example, the arrow pointing from the cylindrically orthotropic material model to the helically orthotropic material model shows that the cylindrically orthotropic material model is a subset of the helically orthotropic material model. (C) Each arrow with a triangular tip(\opentriangle) marks the relationship between a cylindrically homogeneous material model and a Cartesian homogeneous material model. For example, the arrow pointing from the Cartesian transversely isotropic material model to the cylindrically transversely isotropic material model shows that the former is a subset of the latter. }

\label{fig:interdependence}
\end{figure}
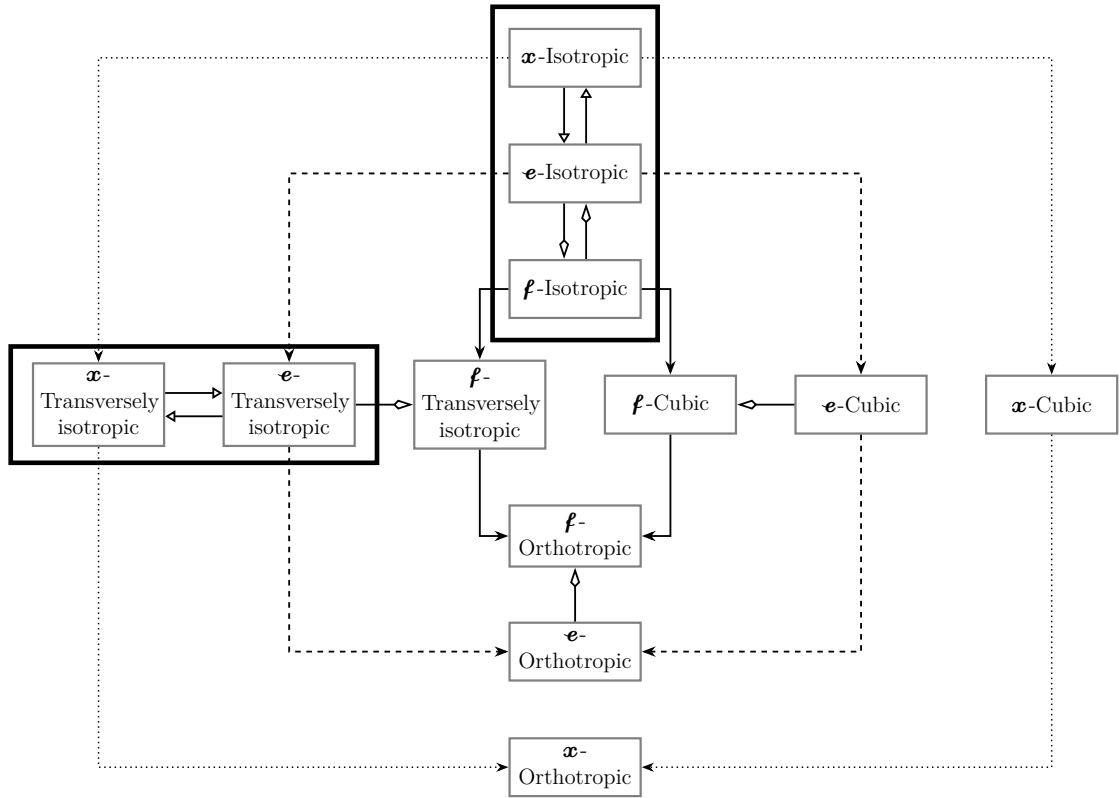

In \S\ref{sec:CylindricallyHelicalSymmetries} we discussed material models in which the material is either orthotropic, transversely isotropic, cubic, or isotropic in the helical, cylindrical, or Cartesian basis. These 12 material models are represented by the 12 boxes in Fig.\ref{fig:interdependence}.
However, it can be shown that if a material is isotropic in one of the three bases, then it is also isotropic in the other two bases. We highlight this identification by drawing a box around the  helically, cylindrically, and Cartesian isotropic material models in Fig.\ref{fig:interdependence}. Also, it can be shown that if a material is transversely isotropic in the cylindrical basis then it is also  transversely isotropic in the Cartesian basis, and \textit{vice versa}. We highlight this identification by drawing a box around the cylindrically and Cartesian transversely isotropic material models in Fig.\ref{fig:interdependence}.
Due to these identifications each of the 12 material models fall into one of following 9 categories.

\begin{enumerate}[label=(\textsc{mm.\roman*}) ,leftmargin=3.5\parindent]
    \item helically orthotropic,
    \item helically transversely isotropic,
    \item  helically cubic,
    \item  cylindrically orthotropic,
    \item  cylindrically cubic,
    \item  helically/cylindrically/Cartesian isotropic,
    \item  cylindrically/Cartesian transversely isotropic,
    \item  Cartesian orthotropic,
    \item  Cartesian cubic.
\end{enumerate}

The above 9 cases, however, are not completely independent of each other.
For example, if a material is cylindrically orthotropic then it is also helically orthotropic.  That is, the set of all cylindrically orthotropic materials is a subset of the set of all helically orthotropic materials.
We highlight this subset relationship  by drawing an  arrow from the box representing cylindrically orthotropic material model to the box representing helically orthotropic material model in Fig.\ref{fig:interdependence}.
We identify all other subset relationships in Fig.\ref{fig:interdependence} by drawing arrows in a similar fashion.

Following the different paths identified by the arrows in Fig.\ref{fig:interdependence} it can be seen that leaving out the Cartesian orthotropic material model (\textsc{mm.viii}) and the Cartesian cubic material model (\textsc{mm.ix}), all other material models are special cases of the helically orthotropic material model (\textsc{mm.i}).

\section{Results}
\label{sec:Results}

\subsection{A structural mechanics model for lamellar spicules}
\label{sec:BendingStiffness}

\begin{figure}[t]
  \centering
  \graphicspath{{figure/}}
   \includegraphics[width=\textwidth]{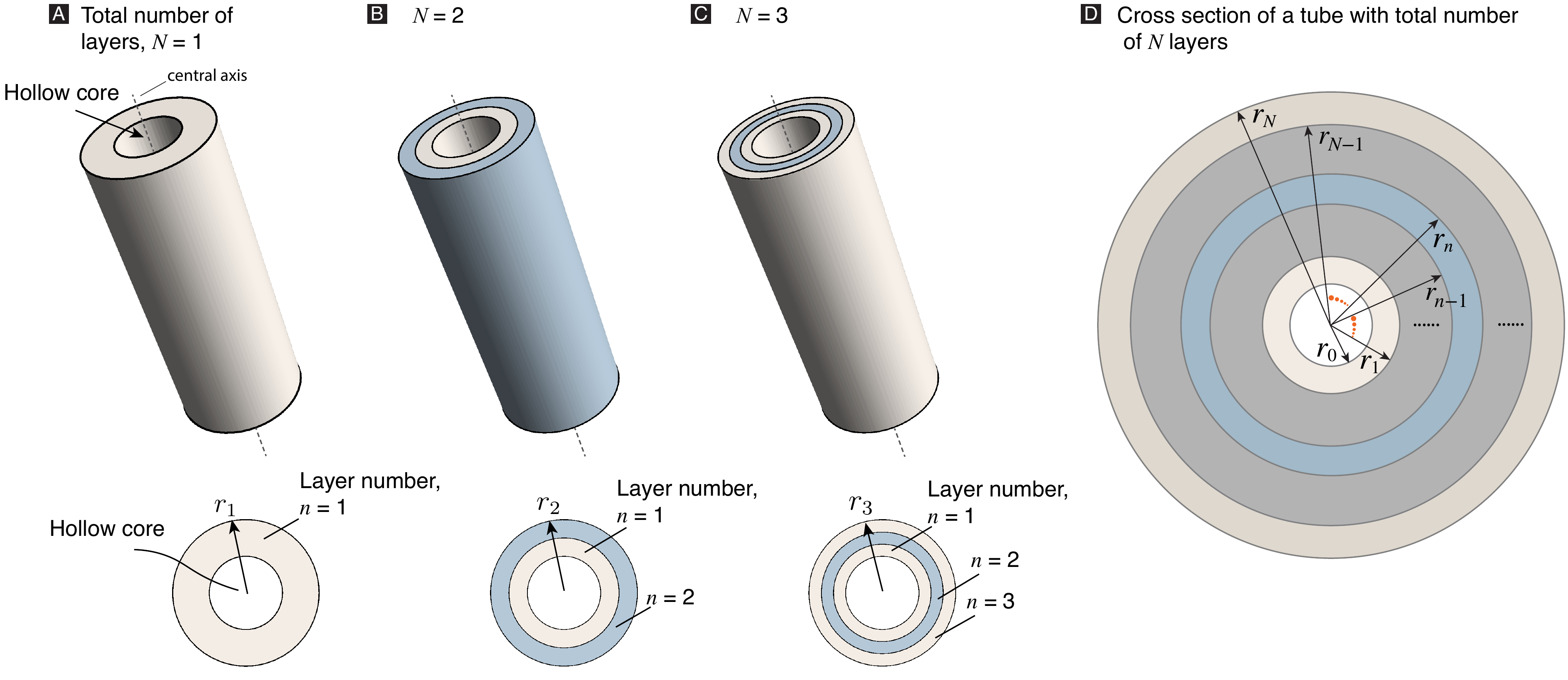}
  \caption{Three-dimensional schematics of  $N$-layer cylindrical assembly.~$\subf{A}$ General view and cross section view for $N = 1$.~$\subf{B}$ General view and cross section view for $N = 2$.~$\subf{C}$ General view and cross section view for $N = 3$.~$\subf{D}$ Cross section view of an arbitrary $N$-layer cylindrical structure. Inner and outer radii for the $1^\text{st}$ layer, $n^\text{th}$ layer and $N^\text{th}$ layer are marked in the figure.}
  \label{fig:Cylinder3D}
\end{figure}

\begin{figure}[t]
  \centering
  \graphicspath{{figure/}}
   \includegraphics[width=0.8\textwidth]{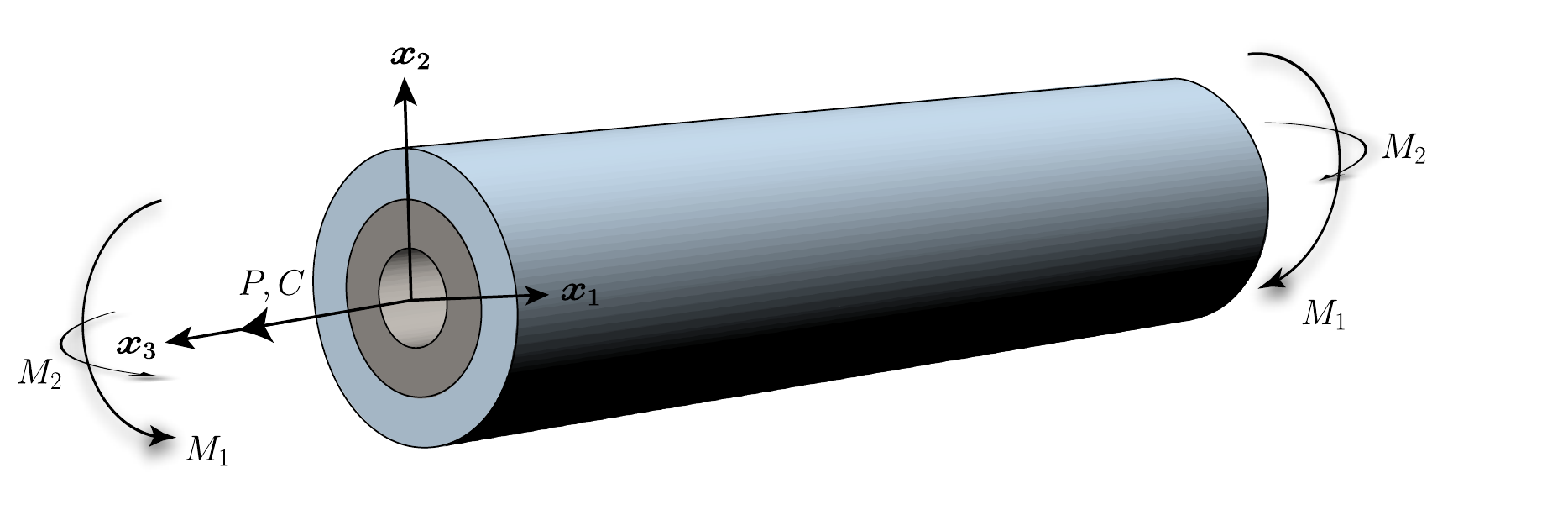}
  \caption{Schematic of the problem studied in Jolicoeur and Cardou~\cite{Jolicoeur1994}. The system is an assembly of cylindrical tubes, made of helically orthotropic materials, that is subjected to an axial force $P$, twisting moment $C$, and bending moments $M_1$, $M_2$. The hypotheses of the problem are:  the strains in the cylinder are small; the axial load ($P$), the moments ($M_1$, $M_2$, and $C$), and the curvature of the assembly's deformed central axis do not vary in the $\physx_3$-direction; there is no resultant shear force on any of the assembly's cross-sections; and the stresses and strains only depend on~$r$ and~$\theta$. Jolicoeur and Cardou use a stress function approach to solve the problem.}
  \label{fig:JCmodel}
\end{figure}

Building on the work of Lekhnitskii~\cite{Lekhnitskii1981}, Jolicoeur and Cardou~\cite{Jolicoeur1994} developed and studied a structural mechanics model for a tight fitting assembly of concentrically arranged tubes. 
The assembly can include a tight fitting solid cylinder at its center. 
Each of the tubes have the shape, position, and orientation of $\mathcal{\kappa}_{\rm R}\ag{\mathcal{B}}$ (see Fig.\ref{fig:cylinderSchematic}). 
We denote the total number of tubes in the assembly as $N\in \mathbb{Z}_{\ge 1}$, and the inner and outer radii of the  $n^{\rm th}$ tube, where $n\in (1,\ldots,N)$, as $r_{n-1}~U[\mathbb{E}_{\rm{R}}]$ and $r_n~U[\mathbb{E}_{\rm{R}}]$, respectively, where $0<r_{n-1}<r_n$\footnote{For example, if $U[\mathbb{E}_{\rm{R}}]$ was $\textsf{centimeters}$, and the inner and outer radii of the $10^{\rm th}$ tube in an assembly of 20 tubes were $10$ and $15$ \textsf{millimeters}, respectively, then in that case $r_{9}=1.0$ and $r_{10}=1.5$.}.
An illustration of the assembly is shown in  Fig.\ref{fig:Cylinder3D}. 
The cross-sections can support bending moments (marked as $ M_1$ and $M_2$ in Fig.\ref{fig:JCmodel}), twisting moment (marked as $C$ in Fig.\ref{fig:JCmodel}), and an axial force ($P$ in Fig.\ref{fig:JCmodel}). 

We will be referring to the tubes also as cylindrical layers, or simply as layers. 
Jolicoeur and Cardou (JC) consider two cases, one in which the internal contacting  surfaces  are able to freely slip with respect to each other, and another in which they can have no relative slip. (In both cases the surfaces do not undergo any relative normal motion.) 

\subsubsection{Particularization of the JC model}
\label{subsec:PaJCModel}
We use a particularized form of the JC model to understand the bending behavior of the spicules that we discussed in \S\ref{sec:intro}. In our particularization we ignore the twisting moment, and the axial force. 
We also ignore the central cylinder, since the spicule's central core is not exactly a solid cylinder, but, rather, closer to being a hollow cylinder; albeit, one in which the inner radius is quite small. The core has a proteneicious filament  running along its length at its center~\cite{weaver2003molecular}. 
We also take the internal  surfaces to be freely slipping. 
Considering the spicules' geometry, internal structure, material properties, and loading, which we discussed in \S\ref{sec:intro}, this particularization is as applicable as JC's complete model for modeling the mechanics of the spicule while at the same time being substantially simpler than it. 

For the application of  the JC model, and our particularization of it, it is necessary that each tube be composed of a helically orthotropic material. 
(The JC model, however, does not apply to all helically orthotropic materials, see \S\ref{sec:admissibility}.)  
We found that if a material is helically homogeneous then it is also  cylindrically  homogeneous. 
For that reason and by the definition of cylindrical  homogeneity (see \S\ref{sec:MaterialSymmetries}), each layer's $\Ce\ag{\cdot}$ is a constant function. 
We denote the constant value of the $n^{\rm th}$ layer's $\Ce\ag{\cdot}$  as  $\Ce[n]$
\footnote{For clarification, we note that the components of $\Ce[n]$ are referred to as the strain coefficients in the work by Lekhnitskii~\cite{Lekhnitskii1981}.}. 
We refer to $\Ce[n]$ as the $n^{\rm th}$ layer's
cylindrical inverse stiffness matrix
\footnote{
We have defined the cylindrical inverse stiffness matrix in \S\ref{sec:LinearElasticity}.
However, in order to get a more intuitive feel for the components of $\Ce[n]$ imagine that we cut out a cube of material of infinitesimal size from the $n^{\rm th}$ layer about the material particle $\mathcal{x}$ such that the cube's normals are aligned with the cylindrical basis vectors at $\mathcal{x}$, i.e., with the $\physe\ag{\mathcal{x}}$ directions. Then if we were to conduct a uniaxial tensile test on that cube by pulling on its faces that were perpendicular to the $\physe_{1}\ag{\mathcal{x}}$ direction, then the Young's modulus we would measure in that test would be  $1/\Ce[n][1][1]~U\ag{\mathbb{F}}/U\ag{\mathbb{E}_{\rm{R}}}^2$.}.

As per our particularization of the JC model, which we will from here on refer to simply as the JC model, a spicule's bending stiffness is related to its layers' thicknesses and material properties as
\begin{subequations}
\begin{align}
\mathcal{K}_{N}\ag{\darg}U\ag{\mathbb{F}} U\ag{\mathbb{E}_{\rm{R}}}^2,
\intertext{where}
\darg=\pr{\pr{r_{n-1},r_n-r_{n-1},C^{\physe}_n}}_{n=\pr{1,\ldots,N}}\label{eq:xarg},
\end{align}
\label{eq:BendingStiffnessAssembly1}
\end{subequations}
and $\mathbb{F}$ denotes the force vector space in our problem.

 The function $\mathcal{K}_N$, which appears in \eqref{eq:BendingStiffnessAssembly1}, is defined as $\mathcal{K}_{N}:
\pr{
\mathbb{R}_{>0},
\mathbb{R}_{>0},
\mathcal{M}_{6 \times 6}\pr{\mathbb{R}}
}^{N}\to \mathbb{R}$,
\begin{equation}
    \mathcal{K}_N\ag{\darg} :=  \sum_{n=1}^{N} \mathcal{k}\ag{\darg_{\cdot n}}.
    \label{eq:BendingStiffnessAssembly2}
\end{equation} 
The function $\mathcal{k}$, which appears in \eqref{eq:BendingStiffnessAssembly2}, is defined as
$\mathcal{k}:\mathbb{R}_{>0}\times \mathbb{R}_{>0}\times \mathcal{M}_{6\times 6}\pr{\mathbb{R}}\to \mathbb{R}$,
\begin{subequations}
\begin{align}
\mathcal{k}\ag{a,t,s}&= \gamma\ag{s}I\ag{a,t,2}+\sum_{i = 1}^{4} \alpha\ag{s}_{\cdot i} K\ag{a,t,s}_{\cdot i}  I\ag{a,t,m\ag{s}_{\cdot i}},\label{eq:kLayer}
\intertext{where $I:\mathbb{R}_{> 0}\times \mathbb{R}_{> 0}\times \mathbb{R}\to \mathbb{R} $,}
I\ag{a,t,m}&=\p\pr{a,m+2}-\p\pr{a+t,m+2}.\label{eq:KIdef}
\end{align}
\label{eq:LayerBendingStiffnessdef}
\end{subequations} In \eqref{eq:BendingStiffnessAssembly2} the argument ``$x_{\cdot n}$'' is to be interpreted as ``the $n^{\rm th}$ component of $x$.'' In general, we use the notation that when $y\in \mathcal{M}_{m\times n}\pr{\mathbb{R}}$, then $y_{\cdot i\cdot j}$, where $1\le i\le m$, $1\le j\le n$, is the $i$-$j^{\rm th}$ component of $y$. 

The functions $\alpha:\mathcal{M}_{6\times 6}\pr{\mathbb{R}}\to \mathbb{R}^4$, $m:\mathcal{M}_{6\times 6}\pr{\mathbb{R}}\to \mathbb{R}^4$, and $\gamma:\mathcal{M}_{6\times 6}\pr{\mathbb{R}}\to \mathbb{R}^4$, which appear in \eqref{eq:kLayer} are, respectively, defined in \S\ref{sec:alphani}, \S\ref{sec:mni}, and \S\ref{sec:gammani}. 
The function $K:\mathbb{R}_{>0}\times \mathbb{R}_{>0}\times \mathcal{M}_{6\times 6}\pr{\mathbb{R}}\to \mathcal{M}_{4\times 1}\pr{\mathbb{R}}$, which appears in \eqref{eq:kLayer}, is defined as
\begin{subequations}
\begin{align}
K\ag{a,t,s}&=\textsf{Inv}\ag{A}b \label{eq:Kdef},
\intertext{where $A\in \mathcal{M}_{4\times 4}\pr{\mathbb{R}}$,}
A&=
\begin{array}{ll}
\bigg(\pr{~\pr{a+t}^{\minusminus{m}_{i}}~|~i\in (1,\ldots,4)},\\
\,\,\,\pr{~a^{\minusminus{m}_{i}}~|~i\in (1,\ldots,4)},\\
\,\,\,\pr{~g_{ i}\pr{a+t}^{\minusminus{m}_{i}}~|~i\in (1,\ldots,4)},\\
\,\,\,\pr{~g_{ i} a^{\minusminus{m}_{i}}~|~i\in (1,\ldots,4)}\bigg),
\end{array}\label{eq:kADef}
\\
m_i&=m\ag{s}_{\cdot i},\quad i\in (1,\ldots, 4),\label{eq:Km}\\ 
g_i&=g\ag{s}_{\cdot i},\quad i\in (1,\ldots, 4), \label{eq:Kg}
\intertext{and $b\in \mathcal{M}_{4\times 1}\pr{\mathbb{R}}$,}
b&=\pr{-\mu_1,-\mu_1,-\mu_2,-\mu_2},\\ 
\mu_i&=\mu\ag{s}_{\cdot i},\quad i\in (1,2). \label{eq:Kmu} 
\end{align}
\label{eq:LayerBendingStiffnessKdefs}
\end{subequations} 

As per our notation, $m\ag{s}_{\cdot i}$ in \eqref{eq:Km} denotes the $i^{\rm th}$ component of $m\ag{s}$, which belongs to $\mathcal{M}_{4\times 1}\pr{\mathbb{R}}$. The symbols $g\ag{s}_{\cdot i}$, and $\mu\ag{s}_{\cdot i}$ appearing, respectively, in \eqref{eq:Kg}, and \eqref{eq:Kmu} are to be interpreted similarly.
As per the new notation we introduce in~\S\ref{Appen:NotationInDePow},
the symbol $\minusminus{m}_i$, which appears in \eqref{eq:kADef}, stands for the expression $m_i-2$. 
The functions $g:\mathcal{M}_{6\times 6}\pr{\mathbb{R}}\to \mathbb{R}^4$, which appears in \eqref{eq:Kg}, and $\mu:\mathcal{M}_{6\times 6}\pr{\mathbb{R}}\to \mathbb{R}^2$, which appears in \eqref{eq:Kmu}, are defined in \S\ref{sec:gni} and \S\ref{sec:muni}, respectively. 

\subsubsection{Admissible material properties}
\label{sec:admissibility}

Jolicoeur and Cardou state that their theory applies whenever a layer is composed of a helically orthotropic material. 

However, we found  that the JC theory, and consequently our particularization of it, does not apply to some special types of helically orthotropic materials.

In \S\ref{sec:CylindricallyHelicalSymmetries} we presented 12 material models, and in \S\ref{sec:mmInterdependence} these models were sorted into 9 categories. These 9 categories are listed as \textsc{mm.i--ix} in \S\ref{sec:mmInterdependence} with \textsc{mm.i} being the helically orthotropic material model. We further identified in  \S\ref{sec:mmInterdependence} that~\textsc{mm.ii}--\textsc{mm.vii} are special cases of \textsc{mm.i}. Within these 7 categories of helically orthtropic material models, we found that the JC theory is inapplicable when the layer is composed of an \textsc{mm.vi} (helically/cylindrically/Cartesian isotropic) material or an \textsc{mm.vii}  (cylindrically/Cartesian transversely isotropic) material.

This restriction that the JC theory cannot be applied to \textsc{mm.vi} and \textsc{mm.vii} materials arises as a consequence of the requirement that for the JC theory to be well-posed the conditions
\begin{subequations}
\begin{align}
m_i&\neq0, \quad i=1,\ 2\label{eq:m1Condition},\\
\left |m_1\right|&\neq\left |m_2\right|,
\label{eq:m2Condition}
\intertext{and}
m_i&\neq2,\quad i=1,\ 2
\label{eq:m3Condition}
\end{align}
\label{eq:mConditions}
\end{subequations}
need to be satisfied.

We refer to the conditions \eqref{eq:mConditions} as the \textit{m}-conditions. 
It is necessary that $(m_1,m_2)$ satisfy \eqref{eq:m1Condition} since otherwise the matrix $A$ (see \eqref{eq:kADef}) would be singular.
For the same reason $(m_1,m_2)$ need to satisfy \eqref{eq:m2Condition}.
The parameters $(m_1,m_2)$ need to satisfy \eqref{eq:m3Condition} since otherwise the $2\times 2$ matrix $B_{\cdot\cdot}$ in \eqref{eq:Bcomponents}, which needs to be inverted to obtain $\mu_i$, will be singular (see \S\ref{sec:Bdeterminant} for details).

It can be shown that \eqref{eq:m3Condition} is violated if a layer is composed of an \textsc{mm.vi} or an \textsc{mm.vii} material.

In summary, for the JC theory and our particularization of it to apply, it is necessary that a layer's material belongs to $\textsc{mm.i}$--$\textsc{mm.v}$. However, belonging to $\textsc{mm.i}$--$\textsc{mm.v}$ may not be sufficient for the JC theory to be applicable. Irrespective, we will only be considering those helically orthotropic materials for which the JC theory is applicable.

\subsection{Sufficient conditions for reduction of bending stiffness}
\label{sec:NecessaryC}

Say an assembly consisting of layers with admissible material properties is subject to a bending moment and as a result attains a curvature of $\kappa_1U\ag{\mathbb{E}_{\rm{R}}}^{-1}$ in the $\physx_1$ direction.
Now keeping the curvature constant, say we were to  create a cylindrical cut in the $n^{\rm th}$ layer of radius $r_c\in \pr{r_{n-1},r_n}$; that is,  make the shear components of the traction vector on the surface $S_{r_c}=\{X\in \kappa_R\ag{\mathcal{B}}~|~\breve{r}\ag{X}=r_c\}$ vanish by allowing the component  of the displacement field parallel to $S_{r_c}$ to be discontinuous across $S_{r_c}$.
If creating such a cut reduces the bending moment on the assembly, then we say that the cut reduces the assembly's bending stiffness.
(\textit{A.1}) We assume the assembly's bending stiffness can be reduced by creating a cut at a  cylindrical surface in the $n^{\rm th}$ layer iff there exist shear stresses on that surface prior to the creation of the cut.

Let us denote the  component in the $\physe_2$ direction of the traction vector on the surface element perpendicular to the $\physe_1$ direction and centered around $X$ as $\tau_{n\cdot1\cdot2}\ag{X}$. Similarly, $\tau_{n\cdot1\cdot3}\ag{X}$ denotes the component in the $\physe_3$ direction. It follows from the JC model that
\begin{subequations}
\label{eq:tauEqs}
\begin{align}
\tau_{n\cdot1\cdot2}\ag{X}&=\kappa\ag{X}f_n\ag{\breve{r}\ag{X}},\label{eq:tau12}\\
\tau_{n\cdot1\cdot3}\ag{X}&= \kappa\ag{X} h_n\ag{\breve{r}\ag{X}},\label{eq:tau13}\\
\intertext{where}
\kappa\ag{X}&=\pr{\kappa_x \cos\ag{\breve{\theta}\ag{X}} + \kappa_y \sin\ag{\breve{\theta}\ag{X}}},\label{eq:def:kappa}\\
f_n\ag{r}&=\pr{- \sum_{i = 1}^{4} K_{n\cdot i} \p\pr{r, \msubm{n}{i}} - \musub{n}{1} r},\label{eq:def:fn}\\
h_n\ag{r}&=\pr{\sum_{i = 1}^{4} \Ksub{n}{i} \gsub{n}{i} \p\pr{r, \msubm{n}{i}} + \musub{n}{2} r},\label{eq:def:hn}\\
K_{n\cdot i}&=K\ag{r_{n-1},r_{n}-r_{n-1},\Ce[n]}_{\cdot i},\label{eq:def:Kni}\\
m_{n\cdot i}&=m\ag{\Ce[n]}_{\cdot i},\label{eq:def:mni}\\
\musub{n}{i}&=\mu\ag{\Ce[n]}_{\cdot i}.\label{eq:def:muni}
\end{align}
\end{subequations}

We found that when
\begin{equation}
\Ce[n][3][4] \neq 0\lor\Ce[n][1][3] \neq \Ce[n][2][3] ,
 \label{eq:nTC}
\end{equation}
then for almost all $r_c\in (r_{n-1}, r_n)$ creating a cut would decrease the assembly's bending stiffness.  

(\emph{Proposition.1}) In \S\ref{sec:step6} we show that if there exist three or more interior cylindrical surfaces within the $n^{\rm th}$ layer such that $\tau_{n \cdot 1 \cdot 2}$ vanishes then that layer's $\mu_{n\cdot i},~i=1,2$ and $K_{n\cdot i},~i=1,\ldots,4$ are naught. (\emph{Proposition.2}) It follows from \eqref{eq:muni} that when a layer's $\mu_i,~i=1,2$  vanish then $\Ce[n][3][4]=0\land \Ce[n][1][3]=\Ce[n][2][3]$. Taking these two results (\emph{Propositions} \emph{1} and \emph{2})  in conjunction implies that when  \eqref{eq:nTC} holds then there are two or fewer interior cylindrical surfaces within the $n^{\rm th}$ layer where $\tau_{n\cdot 1 \cdot 2}$ vanishes. In other words, when \eqref{eq:nTC} holds there is shear stress at almost all $r$ in $(r_{n-1}, r_n)$. In conjunction with our assumption (\textit{A.1}), this result  implies that when \eqref{eq:nTC} holds creating a cut of an arbitrary radius $r$  in $(r_{n-1}, r_n)$ would almost always lead to decrease in the assembly's bending stiffness. We give other interpretations of our result in \S\ref{sec:Discussion}.

\subsection{Sufficient conditions for there being no bending stiffness reduction with the introduction of a cylindrical cut}
\label{sec:TCsatisfied}

We also found that when 
\begin{align}
    \Ce[n][3][4]=0\land\Ce[n][1][3]=\Ce[n][2][3],
\label{eq:TC}
\end{align}
then the assembly's bending stiffness cannot be reduced by creating a cylindrical cut in the $n^{\rm th}$ layer. The proof of this result is as follows.   The matrix $B_{\cdot\cdot}\ag{\Ce[n]}$, where $B_{\cdot\cdot}\ag{\cdot}$ is defined in \eqref{eq:Bcomponents}, is invertible. As we stated in \S\ref{sec:admissibility}, we only consider those helically orthotropic material properties for which the JC theory remains applicable, and if $B_{\cdot\cdot}\ag{\Ce[n]}$ were not invertible then the JC theory would be inapplicable.
As a consequence of $B_{\cdot\cdot}\ag{\Ce[n]}$  being invertible, when \eqref{eq:TC} holds then it follows from \eqref{eq:muni} that $\mu_{n\cdot 1}=0$ and $\mu_{n\cdot 2}=0$. In fact, when   \eqref{eq:TC} holds in addition to $\mu_{n\cdot 1}$, $\mu_{n\cdot 2}$, the constants $K_{n\cdot i}$ also vanish. This is a consequence of \eqref{eq:LayerBendingStiffnessKdefs} and the matrix $A$ in \eqref{eq:LayerBendingStiffnessKdefs} being invertible. The invertibility of $A$ is  guaranteed for the same reason that was discussed for the invertibility of $B_{\cdot\cdot}\ag{\Ce[n]}$. It follows from \eqref{eq:tau12}--\eqref{eq:def:hn} that when $K_{n\cdot i}$ and $\mu_{n\cdot i}$ vanish then there are no shear stresses on any cylindrical surfaces within the $n^{\rm th}$ layer. This result in conjunction with our assumption (\textit{A.1}) implies that when \eqref{eq:TC} holds then the assembly's bending stiffness cannot be reduced by creating a cylindrical cut anywhere in the $n^{\rm th}$ layer.

\color{black}

\subsection{Particularization to Specific Material Models}
\label{subsec:particularization}

The primary result of this paper is \eqref{eq:nTC}, which gives sufficiency conditions for when layering a cylindrical tube would decrease its bending stiffness. Recall that we stated in \S\ref{sec:admissibility} that the JC theory and our particularization of it only applies to  \textsc{mm.i--v}. Therefore, the result \eqref{eq:nTC} also strictly  only applies to   \textsc{mm.i--v}.  
In order to check the condition in \eqref{eq:nTC} for a particular material we need to know that material's $\Ce[n]$. However, the elastic properties of \textsc{mm.i--v} are usually reported in terms of the sets of elastic constants that we discussed in \S\ref{sec:CylindricallyHelicalSymmetries} and not in terms of $\Ce[n]$'s components. For example, the elastic properties of a helically transversely isotropic material (\textsc{mm.ii}) are given in terms of $E_p$, $E_t$, $\nu_{p}$, $\nu_{pt}$, and $\mu_t$.
For that reason in the remainder of this section we will express \eqref{eq:nTC} for each of the material models  \textsc{mm.i--v} in terms of the set of elastic constants that we introduced for them in \S\ref{sec:CylindricallyHelicalSymmetries}.

\subsubsection{ {\footnotesize(MM.I)}  Helically orthotropic material} 
\label{sec:HelOrthoConditions}
Let the assembly's $n^{\rm th}$ layer be composed of a helically orthotropic material.
The helical inverse stiffness matrix of a helically orthotropic material in terms of the elastic constants $\acute{M}$ is the matrix given in \eqref{eq:Ortho}, which we refer to as  $C^{\acute{M}\pr{\physf_{\varphi}}\veryshortarrow \physf_{\varphi}}$.
Using $C^{\acute{M}\pr{\physf_{\varphi}}\veryshortarrow \physf_{\varphi}}$ and following the process detailed in \S\ref{sec:toeConversion} we can obtain the cylindrical inverse stiffness matrix of a helically orthotropic material in terms of $\acute{M}$. We refer to that matrix as $C^{\acute{M}\pr{\physf_{\varphi}}\veryshortarrow \physe}$.
On substituting $\Ce[n][i][j]$ in \eqref{eq:nTC} with the $i-j$ component of $C^{\acute{M}\pr{\physf_{\varphi}}\veryshortarrow \physe}$ and simplifying we get that the sufficient condition for a reduction in the assembly's bending stiffness on introducing a cut in  the $n^{\rm th}$ layer is \eqref{eq:mm1a}$\lor$\eqref{eq:mm1b}, where \eqref{eq:mm1a} and \eqref{eq:mm1b} are, respectively,
\begin{subequations}
\begin{align}
\frac{2\cos^2{\varphi}\sin{\varphi}}{E_3}-\frac{2\cos{\varphi}\sin^3{\varphi}}{E_2}-\frac{1}{4}\left(\frac{1}{\mu_{23}}-\frac{2\nu_{23}}{E_2}\right)\sin{4\varphi}&\neq 0 \label{eq:mm1a}
\intertext{and}
\begin{split}
-\frac{4\nu_{13}\cos^2{\varphi}}{E_1}+\frac{\nu_{23}(3+\cos{4\varphi})}{E_2}-\frac{4\nu_{12}\sin^2{\varphi}}{E_1}-\left(\frac{1}{E_2}+\frac{1}{E_3}-\frac{1}{\mu_{23}}\right)\sin^2{2\varphi}&\neq0\label{eq:mm1b}.
\end{split}
\end{align}
\label{eq:HelOrthoConditions}
\end{subequations} 

\subsubsection{{\footnotesize(MM.II)} Helically transversely isotropic material} 

Let the assembly's $n^{\rm th}$ layer be composed of a helically transversely isotropic material.
The helical inverse stiffness matrix of a helically transversely isotropic material in terms of the elastic constants $\tilde{M}$ is the matrix given in \eqref{eq:TISO}, which we refer to as  $C^{\tilde{M}\pr{\physf_{\varphi}}\veryshortarrow \physf_{\varphi}}$.
 Using $C^{\tilde{M}\pr{\physf_{\varphi}}\veryshortarrow \physf_{\varphi}}$ and following the process detailed in \S\ref{sec:toeConversion} we can obtain the cylindrical inverse stiffness matrix of a helically transversely isotropic material in terms of $\tilde{M}$.
 We refer to that matrix as $C^{\tilde{M}\pr{\physf_{\varphi}}\veryshortarrow \physe}$.
 Substituting $\Ce[n][i][j]$ in \eqref{eq:nTC} with the $i-j$ component of $C^{\tilde{M}\pr{\physf_{\varphi}}\veryshortarrow \physe}$ and simplifying we get that the sufficient condition for a reduction in the assembly's bending stiffness on introducing a cut in  the $n^{\rm th}$ layer is \eqref{eq:mm2a}$\lor$\eqref{eq:mm2b}, where \eqref{eq:mm2a} and \eqref{eq:mm2b} are, respectively, 
\begin{subequations}
\begin{align}
\frac{2\cos^3{\varphi}\sin{\varphi}}{E_t}-\frac{2\cos{\varphi}\sin^3{\varphi}}{E_p}-\frac{1}{4}\left(\frac{1}{\mu_t}-\frac{2\nu_{pt}}{E_p}\right)\sin{4\varphi}&\neq 0\label{eq:mm2a}
\intertext{and}
\begin{split}
-\frac{((-E_p E_t+(E_p+E_t)\mu_t)\cos^2{\varphi}+E_t \mu_t(\nu_p+\nu_{pt}\cos{2\varphi}))\sin^2{\varphi}}{E_p E_t \mu_t}&\neq0\label{eq:mm2b}.
\end{split}
\end{align}
\label{eq:HelTransIsoConditions}
\end{subequations}

\subsubsection{{\footnotesize(MM.III)} Helically cubic material} 
Let the assembly's $n^{\rm th}$ layer be composed of a helically cubic material.
The helical inverse stiffness matrix of a helically cubic material in terms of the elastic constants $\grave{M}$ is the matrix given in \eqref{eq:cubic}, which we refer to as  $C^{\grave{M}\pr{\physf_{\varphi}}\veryshortarrow \physf_{\varphi}}$. 
Using $C^{\grave{M}\pr{\physf_{\varphi}}\veryshortarrow \physf_{\varphi}}$ and following the process detailed in \S\ref{sec:toeConversion} we can obtain the cylindrical inverse stiffness matrix of a helically cubic material in terms of $\grave{M}$.
We refer to that matrix as $C^{\grave{M}\pr{\physf_{\varphi}}\veryshortarrow \physe}$.
Substituting $\Ce[n][i][j]$ in \eqref{eq:nTC} with the $i-j$ component of $C^{\grave{M}\pr{\physf_{\varphi}}\veryshortarrow \physe}$ and simplifying we get that the sufficient condition for a reduction in the assembly's bending stiffness on introducing a cut in  the $n^{\rm th}$ layer is 
\eqref{eq:mm3a}$\lor$\eqref{eq:mm3b}, where \eqref{eq:mm3a} and \eqref{eq:mm3b} are, respectively, 
\begin{subequations}
\begin{align}
\left(\frac{2}{E_c}-\frac{1}{\mu_c}+\frac{2\nu_c}{E_c}\right)\sin (4 \varphi )\neq 0,\label{eq:mm3a}
\intertext{and}
\left(\frac{2}{E_c}-\frac{1}{\mu_c}+\frac{2\nu_c}{E_c}\right)\cos^2{\varphi}\sin^2{\varphi}\neq 0\label{eq:mm3b}.
\end{align}
\label{eq:HelCubicConditions}
\end{subequations}

\subsubsection{{\footnotesize(MM.IV)} Cylindrically orthotropic material}
Let the assembly's $n^{\rm th}$ layer be composed of a cylindrically orthotropic material.
The cylindrical inverse stiffness matrix of a cylindrically orthotropic material in terms of the elastic constants $\acute{M}$ is the matrix given in \eqref{eq:Ortho}, which we refer to as  $C^{\acute{M}\pr{\physe}\veryshortarrow \physe}$. Substituting $\Ce[n][i][j]$ in \eqref{eq:nTC} with the $i-j$ component of $C^{\acute{M}\pr{\physe}\veryshortarrow \physe}$, we get that the sufficient condition for a reduction in the assembly's bending stiffness is $0\neq 0$$\lor$\eqref{eq:CylOrthoConditions}, where \eqref{eq:CylOrthoConditions} is given by
\begin{align}
    \frac{\nu_{13}}{E_1}\neq  \frac{\nu_{23}}{E_2}.
\label{eq:CylOrthoConditions}
\end{align}
Since of course the condition $0\neq0$ can never be satisfied, in the case of cylindrically orthotropic materials the sufficient condition \eqref{eq:nTC} reduces to \eqref{eq:CylOrthoConditions}.

\subsubsection{{\footnotesize(MM.V)} Cylindrically cubic material}

The condition \eqref{eq:nTC} is never satisfied. We arrive at this result in the following manner.
Let the assembly's $n^{\rm th}$ layer be composed of a cylindrically cubic material. The cylindrical inverse stiffness matrix of a cylindrically cubic material in terms of the elastic constants $\grave{M}$ is the matrix given in \eqref{eq:cubic}, which we refer to as  $C^{\grave{M}\pr{\physe}\veryshortarrow \physe}$. 
Substituting $\Ce[n][i][j]$ in \eqref{eq:nTC} with the $i-j$ component of $C^{\grave{M}\pr{\physe}\veryshortarrow \physe}$ we get that the sufficient condition for a reduction in the assembly's bending stiffness on introducing a cut in  the $n^{\rm th}$ layer is  $0\neq 0\lor-\frac{\nu_{c}}{E_c}\neq -\frac{\nu_{c}}{E_c}$, which, of course, can never be satisfied. 

That \eqref{eq:nTC} is never true is equivalent to saying that $\Ce[n][3][4]=0\land\Ce[n][1][3]=\Ce[n][2][3]$. As discussed in \S\ref{sec:TCsatisfied}, this implies that for the of case of a layer composed of  a cylindrically cubic material there will be no  reduction in bending stiffness on introducing a cut in that layer.

\section{Discussion}
\label{sec:Discussion}

Using the equations given in \S\ref{subsec:PaJCModel},
we  numerically computed the bending stiffnesses of several artificial assemblies that contained different number of layers and different material models.
We checked the theoretical results that we presented in \S\ref{sec:NecessaryC} and \S\ref{sec:TCsatisfied} by comparing the bending stiffnesses of two assemblies, where one of them could be considered to have been obtained by  introducing a cylindrical cut in one of the layers of the other assembly. In all our comparisons we found the bending stiffnesses of the two assemblies to be consistent with the results we presented in \S\ref{sec:NecessaryC} and \S\ref{sec:TCsatisfied}.  We discuss a few of the representative calculations that we undertook to check our results in \S\ref{sec:BendingStiffnessReduction}.

To be consistent with the result presented in \S\ref{sec:NecessaryC} it is only required that when \eqref{eq:nTC} is satisfied in a layer the assembly's bending stiffness decrease for "almost all" choices of the cylindrical cut's radius in that layer. A more concrete interpretation of our result in \S\ref{sec:NecessaryC} is that when a layer satisfies \eqref{eq:nTC} and if the bending stiffness of the assembly does not decrease with the first or the second choice of the cylindrical surface  to make the cut in the layer, then it must decrease with the third choice for  the cylindrical cut in the layer. 
Alternatively, our result in \S\ref{sec:NecessaryC} can also be interpreted to mean that if an assembly's bending stiffness does not decrease with the introduction of first two cylindrical cuts in a layer that satisfies \eqref{eq:nTC}, then
the bending stiffness is guaranteed to decrease with the introduction of an additional third cut.
A final interpretation can be that creating three simultaneous cylindrical cuts in a layer satisfying \eqref{eq:nTC} guarantees a decrease in the assembly's bending stiffness. 
In all our numerical calculations we found that the bending stiffness decreased with our very first choice for the cylindrical surface to make the cut.  

To be consistent with the result presented in \S\ref{sec:TCsatisfied} it is only required that when the elastic constants of a layer satisfy the condition \eqref{eq:TC} then introducing a cylindrical cut in that layer does not reduce the assembly's bending stiffness. In all our calculations we found this, in fact, to be the case.

\subsection{Bending Stiffness Reduction}
\label{sec:BendingStiffnessReduction}

\begin{figure}[H]
    \centering
    \graphicspath{{figure/}}
    \includegraphics[width=1.0\textwidth]{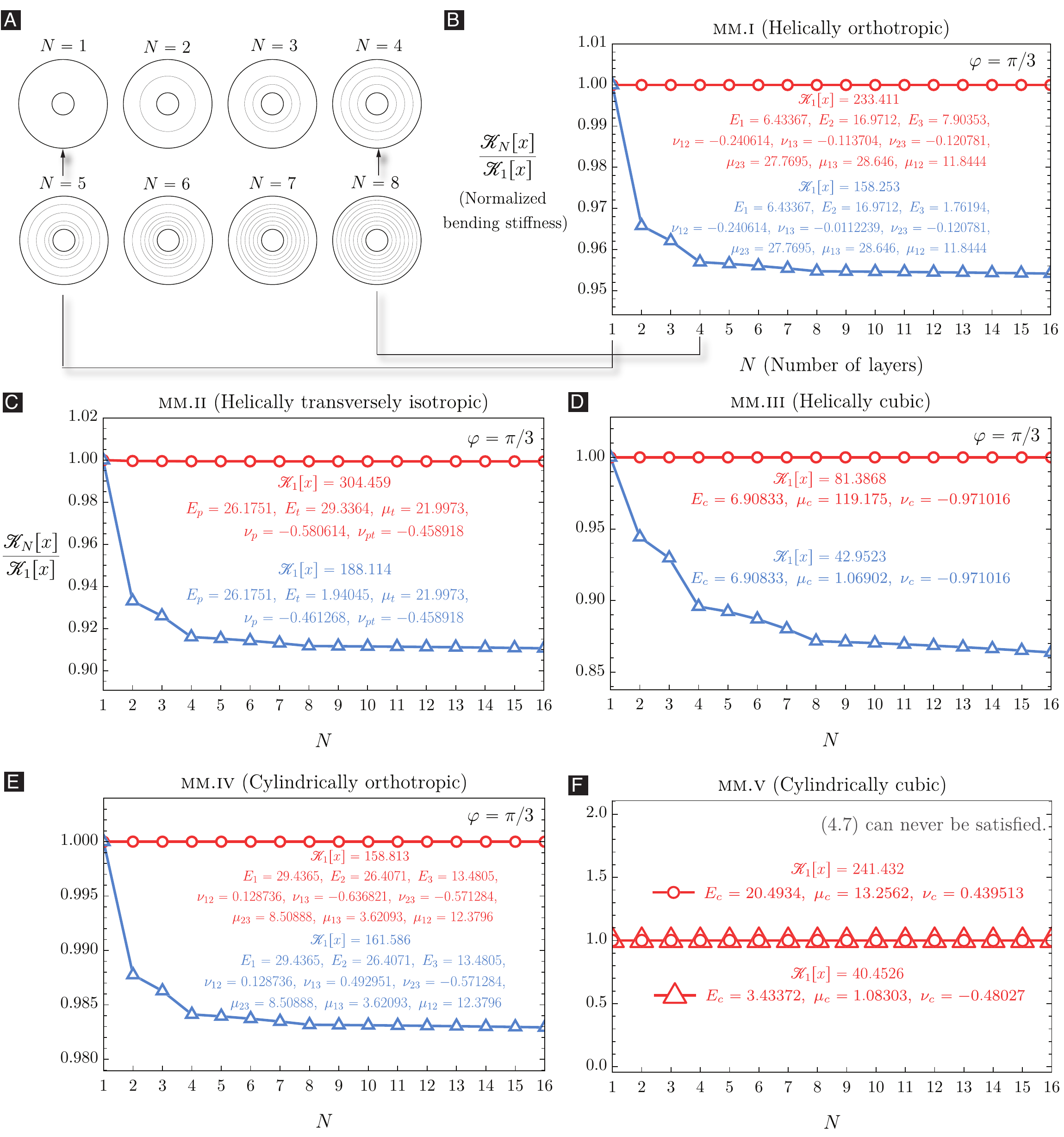} 
    \caption{Normalized bending stiffness, $\mathcal{K}_N\ag{x}/\mathcal{K}_1\ag{x}$, as a function of number of layers, $N$. 
    Subfigures $\subf{B}$--$\subf{F}$ correspond to the material models \textsc{mm.i}--\textsc{mm.v} in this order.
    We computed the quantity $\mathcal{K}_N\ag{x}$ using \eqref{eq:LayerBendingStiffnessKdefs}. For constructing the argument $x$ we need the information about the assembly's layers' radii, and each layer's cylindrical inverse stiffness matrix, $\Ce[n]$. 
     Each $N$-layer-assembly has inner and outer radii of 1 $U\ag{\mathbb{E}_{\rm R}}$ and 2 $U\ag{\mathbb{E}_{\rm R}}$, respectively, and can be thought of as having been obtained by taking an assembly containing $N-1$ layers and creating a new cylindrical cut in the manner described in \S\ref{sec:BendingStiffnessReduction} (see subfigure $\subf{A}$ for illustration). 
    For layers consisting of \textsc{mm.i} or \textsc{mm.iv} we prescribe their elasticity by prescribing numerical values for the elastic constants $\acute{M}$; for layers consisting of \textsc{mm.ii} by prescribing numerical values for the elastic constants $\tilde{M}$; and for layers consisting of \textsc{mm.iii} or \textsc{mm.v} by prescribing numerical values for $\grave{M}$ (cf.  \S\ref{sec:MaterialSymmetries}). We consider two sets of elastic constants for each material model, and their corresponding results are shown using circular and triangular markers, respectively; the red color corresponds to the case where the elastic constants satisfy \eqref{eq:TC}, while the blue color corresponds to the case where the elastic constants satisfy \eqref{eq:nTC}.
    We generated the numerical values for the elastic constants randomly under the constraint that all the corresponding stiffness matrices be positive definite. 
    We list the selected numerical values for the elastic constants for the material models \textsc{mm.i}--\textsc{v} in their corresponding subfigures, i.e., in subfigures $\subf{B}$--$\subf{F}$, respectively. 
    We selected the value of $\pi/3$ for the  helical angle in the material models \textsc{mm.i}--\textsc{mm.iii}. Since we create an $N+1$-layer-assembly by creating a cut in an $N$-layer-assembly, all layers in an assembly (in fact all layers from  all the assemblies  that correspond to the same set of  elastic constants in a subfigure) have the same  cylindrical inverse stiffness matrix.  
    For the material model \textsc{mm.iv} (resp. \textsc{mm.v}) we computed the cylindrical inverse stiffness matrix  using the prescribed numerical $\acute{M}$ (resp. $\grave{M}$) values and \eqref{eq:Ortho} (resp. \eqref{eq:cubic}).
    For the material model \textsc{mm.i} (resp. \textsc{mm.ii},  \textsc{mm.iii}) we first computed the helical inverse stiffness matrix $C^{\acute{M}(\physf_{\varphi})\veryshortarrow \physf_{\varphi}}$ $\left(\text{resp.}\  C^{\tilde{M}(\physf_{\varphi})\veryshortarrow \physf_{\varphi}},\ C^{\grave{M}(\physf_{\varphi})\veryshortarrow \physf_{\varphi}}\right)$  using the prescribed numerical $\acute{M}$ (resp. $\tilde{M}$, $\grave{M}$) values and \eqref{eq:Ortho} (resp. \eqref{eq:TISO}, \eqref{eq:cubic}), and from that  obtained the cylindrical inverse stiffness matrix $C^{\acute{M}(\physf_{\varphi})\veryshortarrow \physe}$ $\left(\text{resp.}\ C^{\tilde{M}(\physf_{\varphi})\veryshortarrow \physe},\ C^{\grave{M}(\physf_{\varphi})\veryshortarrow \physe}\right)$ using the procedure described in \S\ref{sec:toeConversion}.}
    \label{fig:ReductionPlot}
\end{figure}

Figure \ref{fig:ReductionPlot} shows the bending stiffnesses of assemblies in which all the layers are composed of the same helically orthotropic material. The bending stiffnesses of  assemblies composed of \textsc{mm.i--v} are, respectively, shown in  subfigures $\subf{B}$--$\subf{F}$.

In all the assemblies the inner and out radii, $r_0$ and $r_N$, were taken to be 1 $U\ag{\mathbb{E}_{\rm R}}$ and 2 $U\ag{\mathbb{E}_{\rm R}}$, respectively.
We created the geometry of an assembly of $N+1$ layers  by taking an assembly containing $N$ layers and  cutting its $k^{th}$ layer into two layers of equal thickness, where 
\begin{align}
k=2&\pr{N-2^{\lfloor\log_2\ag{N}\rfloor}}+1,
\end{align}
and $\lfloor\cdot\rfloor$ is the floor function.
As a consequence of creating each assembly's geometry in this fashion, the assembly containing $N$ layers can always be thought of as having been created by introducing a cylindrical cut in an assembly containing $N-1$ layers (see Fig.\ref{fig:ReductionPlot}$\subf{A}$). Therefore the validity of the results presented in \S\ref{sec:NecessaryC} and \S\ref{sec:TCsatisfied} can be checked by directly comparing the bending stiffnesses of the $N$ and $N-1$ layer assemblies.

In the subfigures of Fig.\ref{fig:ReductionPlot} each point corresponds to a different assembly, with the abscissa and ordinate of the point providing the assembly's number of layers, $N$, and the assembly's normalized bending stiffness, respectively. The normalized bending stiffness of an assembly is the quantity
$\mathcal{K}_{N}\ag{x}/\mathcal{K}_1\ag{x}$, where,  
recall that, $\mathcal{K}_N\ag{x}$ is the $N$-layer-assembly's non-dimensional bending stiffness  and $x$ encapsulates the assembly's internal geometry and material property information (see \eqref{eq:xarg}).

Recall that the subfigures $\subf{B}$--$\subf{F}$ correspond to the material models \textsc{mm.i--v}, respectively.
Leaving  out \textsc{mm.v} for each material model we consider two sets of elastic constants: one for which \eqref{eq:nTC} is satisfied, and the other for which  \eqref{eq:TC} is satisfied. For \textsc{mm.v} we consider two sets of elastic constants both of which satisfy \eqref{eq:TC}, since  the elastic constants for this material model can never satisfy \eqref{eq:nTC}.

In each subfigure, the results for the set of elastic constants for which \eqref{eq:nTC} is satisfied are shown in blue, whereas the results for the set of elastic constants for which  \eqref{eq:TC} is satisfied are shown in red. For each material model we give the values for the two sets of elastic constants used in the corresponding subfigure. The elastic constants that we use for each material model are the ones that we introduced for them in \S\ref{sec:MaterialSymmetries}. Calculation of $\mathcal{K}_N\ag{x}$ requires calculating the $\Ce[n]$ from the given elastic constant values. The procedure for doing so is detailed in the caption of Fig.\ref{fig:ReductionPlot}.

 From the relative position of the blue points in each subplot it can be noted that when \eqref{eq:nTC} was satisfied the creation of a cut has always led to a decrease in the bending stiffness. This observation is consistent with the  theory presented in \S\ref{sec:NecessaryC}. For an elaboration on some of the different ways of interpreting the result presented in \S\ref{sec:NecessaryC} see the preamble of \S\ref{sec:Discussion}.

From the relative position of the red points in each subplot it can be noted that when \eqref{eq:TC} was satisfied the creation of a cut never led to a decrease in the bending stiffness. This observation is in perfect accordance with the result we presented in  \S\ref{sec:TCsatisfied}, where we stated that when  \eqref{eq:TC} is satisfied then the introduction of a cut should not lead to a reduction in bending stiffness.

\section{Concluding remarks}
\label{sec:conclusion}

In this section we collect and list the most pertinent assumptions in our work, so that it is easier to ascertain the validity of our arguments and results in the future.  

(A) By focusing on the LCC  in the looped configuration we are assuming that most spicules function in the looped configuration within the ocean sediment, rather than in a straight configuration. Future experimental studies, such as that using X-ray computed tomography, would be useful for establishing the validity of this assumption.

(B) We stated that the spicule's LCC in a looped configuration inversely depends on its effective bending stiffness. This statement is not strictly an assumption, since it is based on our study of a nonlinear, small strains and large rotations, beam model for the spicule. We list it here since it is a critical part of our arguments in support of the LCC hypothesis. 

(C) In formulating our structural mechanics model for the spicule we assumed that the spicule's silica layers freely slip with respect to each other. We based this assumption on AFM experiments~\cite{weaver2007hierarchical} that reveal that the protein interlayers are highly compliant compared to the silica layers. However, a more direct measurement of the shear stresses being transmitted between the silica layers would help ascertain the validity of this assumption.

(D)  Our deductions in \S\ref{sec:NecessaryC} and \S\ref{sec:TCsatisfied} are based on the key assumption that the assembly's bending stiffness can be reduced by creating a cut at a  cylindrical surface  iff there exist shear stresses on that surface prior to the creation of the cut.  
In small strain and small rotation beam theories this assumption can be checked to be true. 
However, it needs to be checked for the case of  large rotations which the spicules, quite clearly, undergo  during their operation.  

(E) When we state that our work shows that it is possible for the layers in the spicule to reduce its bending stiffness, we are implicitly assuming that it is possible for the spicules to possess the anisotropic elasticity characteristics that make such a reduction possible. It certainly is not unreasonable to make such an assumption, since 
the anisotropic elasticity of the spicules has not been fully characterized and nothing from the growth and formation processes of the spicules precludes the possibility that the spicules  possess the necessary anisotropic elasticity characteristics. Nevertheless, our results also do not preclude the possibility that the layers do not contribute to the spicule's LCC in the manner we discussed in this paper. It is possible that their anisotropic elasticity characteristics do not allow our theory to be applicable (e.g., their elasticity may not be that of a helically orthotropic material). If the spicule's anisotropic elasticity characteristics allow our theory to be applied, then it is possible that their elastic constants satisfy \eqref{eq:TC} (rather than \eqref{eq:nTC}), thus essentially guaranteeing that the spicule's layers do not contribute to its bending stiffness and therefore do not contribute to the spicule's LCC in the manner we suggest in this paper, if at all. Therefore, the primary value of our results is in that they 
make a precise and direct check of the LCC hypothesis possible once the spicule's anisotropic elasticity becomes better characterized.

\section*{Acknowledgements}

H.K. thanks the American Society of Mechanical Engineers and Brown University for the support provided to him through the Haythornthwaite Research Initiation Grant and the Richard B. Salomon Faculty Research Award, respectively.
S.K. and B.G. were partially supported by the Panther II program, and Panther III and TIGER program, respectively.
S.K. is also grateful for the support she received through Hibbitt fellowship at Brown University.

\newpage
\appendix

\section{Notation}
\label{Appen:Notation}

We denote the set of real numbers as $\mathbb{R}$, the set of non-negative real numbers as $\mathbb{R}_{\ge 0}$, the set of positive real numbers as $\mathbb{R}_{>0}$, the set of integers as $\mathbb{Z}$, and the set of natural numbers as $\mathbb{Z}_{\ge 1}$.

The notation discussed in the previous paragraph is quite standard. 
We have attempted to restrict ourselves to only using standard notation. 
However, in order to simplify some of the most cumbersome looking expressions, we decided to introduce and use some new notation. 
We introduce this new notation in \S\ref{Appen:NotationNonDimenDimen}, \S\ref{Appen:NotationInDePow}, and \S\ref{Appen:OrderedSets}.  

\subsection{Non-dimensional and dimensional quantities\label{Appen:NotationNonDimenDimen}} 

All symbols in which the base symbol is a regular typeface latin character represent non-dimensional variables or numbers. 
For example, we may denote a real number as $x$, and a collection of $n\in \mathbb{Z}_{\ge 1}$ real numbers as $x_1$, $x_2,\ldots,x_n$. 
The typefaces of superscripts and subscripts do not in generally carry any special meaning. 
For example, $x_{a}$,  $x_{\boldsymbol{a}}$, and $x^{\mathscr{a}}$ all denote non-dimensional quantities. 

Symbols in which the base symbol is a script typeface latin character denote dimensional scalar quantities. 
For example, the isotropic Young's modulus is denoted as $\physE$. 
Say $\physE= E~\unit{N/m^2}$, where $E\in \mathbb{R}_{>0}$, then, when there is no confusion, we refer to $E$ too as the isotropic Young's modulus. 

\subsection{Increment, decrement, and exponentiation operations\label{Appen:NotationInDePow}} 

Let $x\in \mathbb{R}$. 
We denote the expressions $x+1$ and $x+2$ sometimes as $\plus{x}$ and $\plusplus{x}$, respectively. 
Similarly, we denote the expressions $x-1$ and $x-2$ sometimes as $\minus{x}$ and $\minusminus{x}$, respectively. 
Each circumflex ($\hat{\cdot}$) overhead a (base) symbol denotes an increment operation on the quantity denoted by that symbol along with any of its superscripts and superscripts, and similarly each overhead caron ($\check{\cdot}$)  denotes an decrement operation. 
That is, $\hat{m}_1$ denotes the expression $m_1+1$ and not $(m+1)_1$. 
We stress that this notation only applies when the quantity denoted by the base symbol along with any of its superscripts and subscripts is a real valued variable or a real valued constant. 

We denote $x^2$, the square of $x$, as $\p\pr{x,2}$, and in general $x^a$, where $a\in \mathbb{R}$,  as $\p\pr{x,a}$. 

\subsection{Denoting ordered sets} 
\label{Appen:OrderedSets}

We denote un-ordered sets using braces $\{\cdot,\ldots,\cdot\}$, and ordered sets using parentheses $(\cdot, \ldots,\cdot)$. 
We use square brackets to identify a function's arguments, like in $f\ag{\cdot}$, where $f$ is some generic function. 
When the argument to a function is a single ordered set, say $(\cdot,\ldots,\cdot)$, then instead of writing $f\ag{(\cdot,\ldots,\cdot)}$ we simply write $f\ag{\cdot,\ldots,\cdot}$. 

Say $\mathsf{X}=(\mathsf{x}_1,\ldots, \mathsf{x}_n)$, where $n\in \mathbb{Z}_{\ge 1}$. 
The $i^{\rm th}$ element (component) of the ordered set $\mathsf{X}$ is referred to as  $\left(\mathsf{X}\right)_{i}$ or $\mathsf{X}_{\cdot i}$. 
The $j^{\rm th}$ component of $\mathsf{X}_{\cdot i}$ is written as either $\mathsf{X}_{\cdot i\cdot j}$ or, when we want to be more explicit, as $\left(\mathsf{X}_{\cdot i}\right)_{j}$ or $\left(\pr{\mathsf{X}}_{i}\right)_{j}$. 
We will be abbreviating ordered sets such as $(x_1,x_2,\ldots, x_n)$ and  $(\mathsf{x}_1,\mathsf{x}_2,\ldots, \mathsf{x}_n)$ as $(x_i)_{i\in (1,\ldots,n)}$ and $(\mathsf{x}_i)_{i\in (1,\ldots,n)}$, respectively. 
We will be abbreviating nested ordered sets such as
\begin{align}
\pr{\pr{x_{ij}}_{j\in (1,\ldots,m)}}_{i\in (1,\ldots,n)},
\intertext{where $m\in \mathbb{Z}_{\ge 1}$, as}
\pr{x_{ij}}_{i\in (1,\ldots,n),j\in (1,\ldots,m)},
\end{align}
and when $m=n$ as
$$
\pr{x_{ij}}_{i,j\in(1,\ldots,n)}.
$$ 

We consider a matrix to be an ordered set of elements where all the elements are ordered sets of the same cardinality. Specially, we denote the  ordered set containing $m\in \mathbb{Z}_{\ge 1}$ elements where each element is an ordered set containing $n\in\mathbb{Z}_{\ge 1}$ real numbers as $\mathcal{M}_{m\times n}\pr{\mathbb{R}}$.

\subsection{Voigt notation}
\label{sec:Voigt}

The function $\textsf{voi}:(1,2,3) \times(1,2,3) \to(1,\ldots,6)$ is defined as
\begin{equation}
    \textsf{voi}\ag{i,j}=
    \left\{
    \begin{array}{ll}
    i, & i= j,\\
    4, & (i,j)=(2,3)~\text{or}~(3,2),\\
    5, & (i,j)=(1,3)~\text{or}~(3,1),\\
    6, & (i,j)=(1,2)~\text{or}~(2,1).
    \end{array}
    \right.
\label{eq:voidef}
\end{equation}
The function $\textsf{voi}^{-1}: (1,\ldots,6)\to (1,2,3)\times (1,2,3)$ is defined as
\begin{equation}
    \textsf{voi}^{-1}\ag{I}=
    \left\{
    \begin{array}{ll}
    (I,I), & I\le 3,\\
    (2,3), & I= 4,\\
    (1,3), & I= 5,\\
    (1,2), & I= 6.
    \end{array}
    \right.
\label{eq:voiinvdef}
\end{equation}

\subsection{Logical operators}
We follow the standard notations and use the symbol $\land$ to indicate logical "and", i.e., if we say that $A\land B$ is true then we mean that both $A$ and $B$ are true. Similarly, we use the symbol $\lor$ to denote logical inclusive "or", i.e., if we say that $A\lor B$ is true then we mean that one of the following three cases is true: (i) $A$ is true and $B$ is not true, (ii) $A$ is not true and $B$ is true, and (iii) both $A$ and $B$ are true.

\section{Definitions of various material and micro-architecture dependent constants}
\label{Appen:MatConst}


\subsection{Definition of  $\beta\ag{\cdot}$\label{sec:betanij}} 

The function $\beta:\mathcal{M}_{6\times 6}\pr{\mathbb{R}}\to \mathcal{M}_{6\times 6}\pr{\mathbb{R}}$ is defined as
\begin{equation}
\beta\ag{s} =
\left(
s_{\cdot i\cdot j}-\frac{s_{\cdot i\cdot 3} s_{\cdot 3\cdot j}}{s_{\cdot 3\cdot 3}}\right)_{i,j\in(1,\ldots,6)}.
\label{eq:betanijdef}
\end{equation} 

\subsection{Definitions of $\alpha\ag{\cdot}$, $m\ag{\cdot}$, and $\gamma\ag{\cdot}$\label{sec:alphamgni}} 

\subsubsection{Definition of $\alpha\ag{\cdot}$\label{sec:alphani}} 

The function $\alpha\ag{\cdot}$ is defined as $\alpha:\mathcal{M}_{6\times 6}\pr{\mathbb{R}}\to \mathbb{R}^4$,
\begin{align}
    \alpha\ag{s}_{\cdot i} & = \frac{\pi}{\pr{m\ag{s}_{\cdot i}+2} s_{\cdot 3\cdot 3}}\left(s_{\cdot 1\cdot 3} + \pr{m\ag{s}_{\cdot i}+1}s_{\cdot 2\cdot 3} - s_{\cdot 3\cdot 4}\, g\ag{s}_{\cdot i}\, m\ag{s}_{\cdot i}\right),
\label{eq:alphani}
\end{align}
where $i\in (1,\ldots,4)$. The functions $m\ag{\cdot}$, and $g\ag{\cdot}$, which appear in \eqref{eq:alphani}, are, respectively, defined in \S\ref{sec:mni}, and \S\ref{sec:gni}.

\subsubsection{Definition of $m\ag{\cdot}$ \label{sec:mni}}

The function $m:\mathcal{M}_{6\times 6}\pr{\mathbb{R}}\to \mathbb{R}^4$\footnote{That $m\ag{\cdot}$ is a real-valued function is an assumption. We take this to be a reasonable assumption based on the observation that in all the cases that we have tested numerically, $m\ag{\cdot}$ has yielded real numbers.} is defined as
\begin{subequations}
    \begin{align}
     m\left[s\right]_{\cdot 1} & =  \sqrt{\frac{-h\ar{s} +\sqrt{\p\pr{h\ar{s},2} - 4d\ar{s} l\ar{s} }}{2d\ar{s}}},
     \intertext{where note that, as per the notation we introduced in \S\ref{sec:BendingStiffness}, $\p\pr{h[s],2}$ denotes the square of $h[s]$,}
     m\left[s\right]_{\cdot 2} & =  \sqrt{\frac{-h\ar{s} -\sqrt{\p\pr{h\ar{s},2} - 4d\ar{s} l\ar{s} }}{2d\ar{s}}}, \\
     m\left[s\right]_{\cdot 3} & =  -m\left[s\right]_{\cdot 1} ,
   \intertext{and}
     m\left[s\right]_{\cdot 4} & =  -m\left[s\right]_{\cdot 2} ,
 \end{align}
 \label{eqns:mnis}
\end{subequations}
where
\begin{subequations}
  \begin{align}
     d\ag{s} & :=  \beta\ag{s}_{\cdot 2\cdot 2}\, \beta\ag{s}_{\cdot 4\cdot4} - \p\pr{\beta\ag{s}_{\cdot 2\cdot 4},2}, \label{eq:def:d}\\[1em]
    \begin{split}
     h\ag{s} & := \beta\ag{s}_{\cdot 2\cdot 4}(2\beta\ag{s}_{\cdot 1\cdot 4} + \beta\ag{s}_{\cdot 2\cdot 4} + 2\beta\ag{s}_{\cdot 5\cdot 6}) + \p\pr{\beta\ag{s}_{\cdot 1\cdot 4},2} \\
     & \quad - \beta\ag{s}_{\cdot 4\cdot 4}(\beta\ag{s}_{\cdot 1\cdot 1} + 2\beta\ag{s}_{\cdot 1\cdot 2} + \beta\ag{s}_{\cdot 2\cdot 2} + \beta\ag{s}_{\cdot 6\cdot 6}) - \beta\ag{s}_{\cdot 2\cdot 2} \beta\ag{s}_{\cdot 5\cdot 5},\label{eq:def:h}
    \end{split}
  \intertext{and}
    l[s] & :=  \beta\ag{s}_{\cdot 5\cdot 5}(\beta\ag{s}_{\cdot 1\cdot 1} + 2\beta\ag{s}_{\cdot 1\cdot 2} + \beta\ag{s}_{\cdot 2\cdot 2} + \beta\ag{s}_{\cdot 6\cdot 6})-\p\pr{\beta\ag{s}_{\cdot 5\cdot 6},2}. \label{eq:def:l}
    \end{align}
  \label{eqns:anbncn}
\end{subequations} 
The function $\beta$, which appear in \eqref{eqns:anbncn}, is defined in \S\ref{sec:betanij}.

\subsubsection{Definition of $\gamma\ar{\cdot}$\label{sec:gammani}}

The function $\gamma\ag{\cdot}:\mathcal{M}_{6\times 6}\pr{\mathbb{R}}\to \mathbb{R}$ is defined as
\begin{align}
    \gamma\ag{s} & = \frac{\pi}{4 s_{\cdot 3\cdot 3}} \left( \mu\ag{s}_{\cdot 1} \left(s_{\cdot 1\cdot 3} + 3s_{\cdot 2\cdot 3}\right) - 2\mu\ag{s}_{\cdot 2} s_{\cdot 3\cdot 4} - 1  \right),
\end{align}
where $\mu\ag{\cdot}$ is defined in \S\ref{sec:muni}. 

\subsection{Definitions of $g\ag{\cdot}$ and $\mu\ag{\cdot}$}
\label{sec:gmuni}

\subsubsection{Definition of $g\ag{\cdot}$}
\label{sec:gni}

The function $g\ag{\cdot}$ is defined as $g:\mathcal{M}_{6\times 6}\pr{\mathbb{R}}\to \mathbb{R}^4$,
    \begin{equation}
    g\ag{s}_{\cdot i} = \frac{\beta\ag{s}_{\cdot2\cdot4}\p\pr{m\ag{s}_{\cdot i},2}+(\beta\ag{s}_{\cdot1\cdot4} + \beta\ag{s}_{\cdot2\cdot4})m\ag{s}_{\cdot i} - \beta\ag{s}_{\cdot5\cdot6}}{\beta\ag{s}_{\cdot4\cdot4}\p\pr{m\ag{s}_{\cdot i},2} - \beta\ag{s}_{\cdot5\cdot5}},
\label{eq:gni}
    \end{equation}
where $i\infour$. 
The functions $\beta\ag{\cdot}$ and $m\ag{\cdot}$, which appear in \eqref{eq:gni}, are defined in \S\ref{sec:betanij} and \S\ref{sec:mni}, respectively.

\subsubsection{Definition of $\mu\ag{\cdot}$}
\label{sec:muni}

The function $\mu\ag{\cdot}$ is defined as $\mu:\mathcal{M}_{6\times 6}\pr{\mathbb{R}}\to \mathbb{R}^2$,
    \begin{equation}
    \mu\ag{s} =
    \frac{1}{s_{\cdot 3\cdot 3}}
    \textsf{Inv}\ag{ B_{\cdot \cdot}\textbf{}\ag{s}}
    \begin{bmatrix}
            2s_{\cdot 3\cdot 4} \\ s_{\cdot 1\cdot 3}-s_{\cdot 2\cdot 3}
    \end{bmatrix}.
\label{eq:muni}
\end{equation}
The function $\beta\ag{\cdot}$, which appears in \eqref{eq:muni}, is defined in \S\ref{sec:betanij}. 
The function $B_{\cdot\cdot}\ag{s}$ is defined as $B_{\cdot\cdot}:\mathcal{M}_{6\times 6}\pr{\mathbb{R}}\rightarrow \mathcal{M}_{4\times 4}\pr{\mathbb{R}}$, 
\begin{subequations}
\label{eq:Bcomponents}
\begin{equation}
B_{\cdot\cdot}\ag{s}=
\begin{bmatrix}
    B_{\cdot 1\cdot 1}\ag{s} & B_{\cdot 1\cdot 2}\ag{s} \\
    B_{\cdot 2\cdot 1}\ag{s} & B_{\cdot 2\cdot 2}\ag{s}
\end{bmatrix},
\end{equation}
where its components are
\begin{align}
B_{\cdot1\cdot1}\ag{s}&=-2\beta\ag{s}_{\cdot 1\cdot 4}-6\beta\ag{s}_{\cdot 2\cdot 4}+\beta\ag{s}_{\cdot 5\cdot 6}, \label{eq:B11} \\
B_{\cdot1\cdot2}\ag{s}&=4\beta\ag{s}_{\cdot 4\cdot 4}-\beta\ag{s}_{\cdot 5\cdot 5}, \label{eq:B12} \\
B_{\cdot2\cdot1}\ag{s}&=-\beta\ag{s}_{\cdot 1\cdot 1}-2\beta\ag{s}_{\cdot 1\cdot 2}+3\beta\ag{s}_{\cdot 2\cdot 2}-\beta\ag{s}_{\cdot 6\cdot 6}, \label{eq:B21} \\
B_{\cdot2\cdot2}\ag{s}&=2\beta\ag{s}_{\cdot 1\cdot 4}-2\beta\ag{s}_{\cdot 2\cdot 4}+\beta\ag{s}_{\cdot 5\cdot 6}. \label{eq:B22}
\end{align}
\end{subequations}

\newpage
\section{Relationship between elastic tensors}
\subsection{Link between tensor/matrix components and those in different basis}
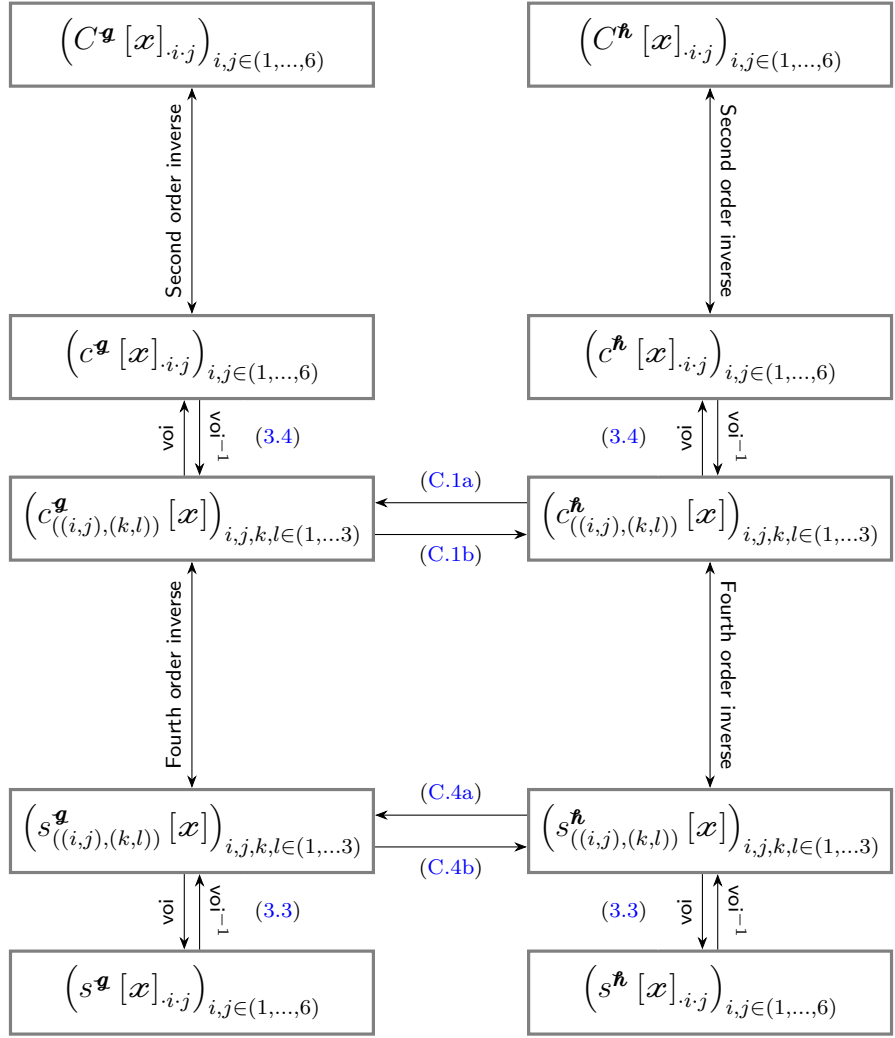
\begin{figure}[!h]
  \centering
      \begin{tikzpicture}[auto,
                     > = Stealth,
         node distance = 5cm and 10cm,
            box/.style = {draw=gray, very thick,
                          minimum height=11mm, text width=45mm,
                          align=center},
     every edge/.style = {draw, <->, very thick},
every edge quotes/.style = {font=\footnotesize, align=center, inner sep=1pt}
                          ]
  \node (n11) [box]               {$\pr{s^{\physga}_{\pr{\pr{i,j},\pr{k,l}}}\ag{\mathcal{x}}}_{i,j,k,l \in (1, \ldots 3)}$};
  \node (n12) [box, below=1cm of n11]{$\pr{s^{\physga}\ag{\mathcal{x}}_{\cdot i \cdot j}}_{i,j\in (1,\ldots, 6)}$};
  \node (n21) [box, above=3cm of n11] {$\pr{c^{\physga}_{\pr{\pr{i,j},\pr{k,l}}}\ag{\mathcal{x}}}_{i,j,k,l \in (1, \ldots 3)}$};
  \node (n22) [box, above=1cm of n21] {$\pr{c^{\physga}\ag{\mathcal{x}}_{\cdot i \cdot j}}_{i,j\in (1,\ldots, 6)}$};
  \node (n13) [box,above=3cm of n22] {$\pr{C^{\physga}\ag{\mathcal{x}}_{\cdot i \cdot j}}_{i,j\in (1,\ldots, 6)}$};

  \node (p11) [box, right=2cm of n11]               {$\pr{s^{\physgb}_{\pr{\pr{i,j},\pr{k,l}}}\ag{\mathcal{x}}}_{i,j,k,l \in (1, \ldots 3)}$};
  \node (p12) [box, below=1cm of p11]{$\pr{s^{\physgb}\ag{\mathcal{x}}_{\cdot i \cdot j}}_{i,j\in (1,\ldots, 6)}$};
  \node (p21) [box, above=3cm of p11] {$\pr{c^{\physgb}_{\pr{\pr{i,j},\pr{k,l}}}\ag{\mathcal{x}}}_{i,j,k,l \in (1, \ldots 3)}$};
  \node (p22) [box, above=1cm of p21] {$\pr{c^{\physgb}\ag{\mathcal{x}}_{\cdot i \cdot j}}_{i,j\in (1,\ldots, 6)}$};
  \node (p13) [box, above=3cm of p22] {$\pr{C^{\physgb}\ag{\mathcal{x}}_{\cdot i \cdot j}}_{i,j\in (1,\ldots, 6)}$};

 \draw[<-] (n11.5) --(p11.175) node[midway, font=\scriptsize, sloped] {\eqref{eq:g1tog2conversion_s}};
  \draw[->] (n11.355) --(p11.185)node[midway,below, font=\scriptsize, sloped] {\eqref{eq:g2tog1conversion_s}};
 \draw[<-] (n21.5) --(p21.175) node[midway, font=\scriptsize, sloped] {\eqref{eq:g1tog2conversion_c}};
 \draw[->] (n21.355) --(p21.185)node[midway,below, font=\scriptsize, sloped] {\eqref{eq:g2tog1conversion_c}};

\draw[<->] (p22)  --   (p13) node[midway, rotate=-180, font=\scriptsize, sloped] {$\textsf{Second order inverse}$};


\draw[->]  (p21.100) -- (p22.260) node[midway,font=\scriptsize, sloped] {$\textsf{voi}$};
\draw[->,white]  (p21.140) -- (p22.220) node[midway,font=\scriptsize,black] {\eqref{eq:cCompVoigt}};
\draw[->,yshift=10mm]  (p22.280) -- (p21.80) node[midway,font=\scriptsize, sloped] {$\textsf{voi}^{-1}$};

\draw[<->]   (p11) -- (p21) node[midway,rotate=-180, font=\scriptsize, sloped] {$\textsf{Fourth order inverse}$};


\draw[<-]  (p12.100) -- (p11.260) node[midway,font=\scriptsize, sloped] {$\textsf{voi}$};
\draw[->,white]  (p12.140) -- (p11.220) node[midway,font=\scriptsize,black] {\eqref{eq:sCompVoigt}};
\draw[<-,yshift=10mm]  (p11.280) -- (p12.80) node[midway,font=\scriptsize, sloped] {$\textsf{voi}^{-1}$};

\draw[<->]   (n11) -- (n21) node[midway, font=\scriptsize, sloped] {$\textsf{Fourth order inverse}$};

\draw[->]  (n21.100) -- (n22.260) node[midway,font=\scriptsize, sloped] {$\textsf{voi}$};
\draw[->,yshift=10mm]  (n22.280) -- (n21.80) node[midway,font=\scriptsize, sloped] {$\textsf{voi}^{-1}$};
\draw[->,white]  (n22.320) -- (n21.40) node[midway,font=\scriptsize,black] {\eqref{eq:cCompVoigt}};


\draw[<->] (n22)  --   (n13) node[midway, font=\scriptsize, sloped] {$\textsf{Second order inverse}$};


\draw[<-]  (n12.100) -- (n11.260) node[midway,font=\scriptsize, sloped] {$\textsf{voi}$};
\draw[<-,yshift=10mm]  (n11.280) -- (n12.80) node[midway,font=\scriptsize, sloped] {$\textsf{voi}^{-1}$};
\draw[->,white]  (n11.320) -- (n12.40) node[midway,font=\scriptsize,black] {\eqref{eq:sCompVoigt}};

      \end{tikzpicture}
  \caption{Procedure for obtaining stiffness, inverse stiffness, and compliance tensors from one another, and also from those in different basis.}
  \label{figure:link}
  \end{figure}

  A fourth order stiffness tensor component in one basis ($\physga\ag{\mathcal{x}}$) is related to that in another basis ($\physgb\ag{\mathcal{x}}$) as
  
  \begin{subequations}
  \begin{equation}
 c^{\physga}_{\pr{(i,j),(k,l)}}\ag{\mathcal{x}}= \sum_{p,q,r,s \incylinder} Q^{\physga \veryshortarrow \physgb}_{\cdot p \cdot i} Q^{\physga \veryshortarrow \physgb}_{\cdot q \cdot j} Q^{\physga \veryshortarrow \physgb}_{\cdot r \cdot k} Q^{\physga \veryshortarrow \physgb}_{\cdot s\cdot l} c^{\physgb}_{\pr{(p,q),(r,s)}}\ag{\mathcal{x}},
 \label{eq:g1tog2conversion_c}
\end{equation}

 \begin{equation}
 c^{\physgb}_{\pr{(i,j),(k,l)}}\ag{\mathcal{x}}= \sum_{p,q,r,s \incylinder} Q^{\physgb \veryshortarrow \physga}_{\cdot p \cdot i} Q^{\physgb \veryshortarrow \physga}_{\cdot q \cdot j} Q^{\physgb \veryshortarrow \physga}_{\cdot r \cdot k} Q^{\physgb \veryshortarrow \physga}_{\cdot s\cdot l} c^{\physga}_{\pr{(p,q),(r,s)}}\ag{\mathcal{x}},
 \label{eq:g2tog1conversion_c}
\end{equation}
\label{eq:cconversion}
\end{subequations}
where $Q^{\physga \veryshortarrow \physgb}_{\cdot \cdot}$ maps $\physga\ag{\mathcal{x}}$ to $\physgb\ag{\mathcal{x}}$ as
\begin{equation}
\physgb_{\dt i}\ag{\mathcal{x}} = \sum_{j \incylinder } Q^{\physga \veryshortarrow \physgb}_{\dt i\dt j}\physga_{\dt j}\ag{\mathcal{x}},
\label{eq:Qg1tog2}
\end{equation}
and $Q^{\physgb \veryshortarrow \physga}_{\cdot \cdot}$, which maps $\physgb\ag{\mathcal{x}}$ to $\physga\ag{\mathcal{x}}$, is given by
\begin{equation}
Q^{\physgb \veryshortarrow \physga} = \textsf{Inv}\ag{Q^{\physga \veryshortarrow \physgb}}.
\end{equation}

Similarly, a fourth order compliance tensor component in one basis ($\physga$) is related to that in another basis ($\physgb$) as
\begin{subequations}
  \begin{equation}
 s^{\physga}_{\pr{(i,j),(k,l)}}\ag{\mathcal{x}}= \sum_{p,q,r,s \incylinder} Q^{\physga \veryshortarrow \physgb}_{\cdot p \cdot i} Q^{\physga \veryshortarrow \physgb}_{\cdot q \cdot j} Q^{\physga \veryshortarrow \physgb}_{\cdot r \cdot k} Q^{\physga \veryshortarrow \physgb}_{\cdot s\cdot l} s^{\physgb}_{\pr{(p,q),(r,s)}}\ag{\mathcal{x}},
 \label{eq:g1tog2conversion_s}
\end{equation}

 \begin{equation}
 s^{\physgb}_{\pr{(i,j),(k,l)}}\ag{\mathcal{x}}= \sum_{p,q,r,s \incylinder} Q^{\physgb \veryshortarrow \physga}_{\cdot p \cdot i} Q^{\physgb \veryshortarrow \physga}_{\cdot q \cdot j} Q^{\physgb \veryshortarrow \physga}_{\cdot r \cdot k} Q^{\physgb \veryshortarrow \physga}_{\cdot s\cdot l} s^{\physga}_{\pr{(p,q),(r,s)}}\ag{\mathcal{x}}.
 \label{eq:g2tog1conversion_s}
\end{equation}
\end{subequations}

\subsection{Obtaining \texorpdfstring{$C^{M\pr{\physg}\veryshortarrow\physe}$}{} from \texorpdfstring{$C^{M\pr{\physg}\veryshortarrow\physg}$}{}}
\label{sec:toeConversion}
In this section we employ the following notation.
We use the symbols $\acute{M}$, $\tilde{M}$, $\grave{M}$, $\bar{M}$ in the superscript to refer to the properties of a orthotropic, transversely isotropic, cubic, and isotropic material respectively.
We use the symbol $\acute{M}\pr{\u{g}}$ in the superscript of a quantity to denote that that quantity corresponds to a general $\u{g}$-orthotropic material. That is, 
the $\physg$ in $\acute{M}\pr{\u{g}}$ can be $\physe$, $\physf_{\varphi}$, or $\physx$.
Similarly, the symbols $\tilde{M}\pr{\u{g}}$, $\grave{M}\pr{\u{g}}$, and $\bar{M}\pr{\u{g}}$ in a quantity's superscript denote that that quantity respectively correspond to a $\u{g}$-transversely-isotropic, $\u{g}$-cubic, and $\u{g}$-isotropic material.
For example, $\mathbbm{c}^{\acute{M}\pr{\physx}}\ag{\mathcal{x}}$ is the elastic stiffness tensor of a Cartesian orthotropic material at the material particle $\mathcal{x}$.
A superscript of $M\pr{\physg}$ denotes an arbitrary material of the type we considered in \S\ref{sec:MaterialSymmetries}. 
That is the $M$ in $M\pr{\physg}$ can stand for $\acute{M}$, $\tilde{M}$, $\grave{M}$, or $\bar{M}$.
The components of the stiffness and compliance tensors can be expressed  w.r.t different bases.
Say that the components are expressed w.r.t the $\physe\ag{\mathcal{x}}$ basis then
we encapsulate that information by appending the superscript of the components, and the superscripts of collections of those components, with the symbol $\veryshortarrow \physe$.
For example, $c^{\acute{M}\pr{\physx}\veryshortarrow \physe}_{\pr{\pr{i,j},\pr{k,l}}}$ denote the components of $\mathbbm{c}^{\acute{M}\pr{\physx}}\ag{\mathcal{x}}$ w.r.t the $\physe\ag{\mathcal{x}}$ basis, and $C^{\tilde{M}\pr{\physf_{\varphi}}\veryshortarrow \physe}$ denotes the inverse stiffness matrix of a helically transversely isotropic material w.r.t the $\physe\ag{\mathcal{x}}$ basis.

Consider an inverse stiffness matrix $C^{M\pr{\physg}\veryshortarrow \physg}$. The matrix $C^{M\pr{\physg}\veryshortarrow \physe}$ can be obtained from $C^{M\pr{\physg}\veryshortarrow \physg}$ by carrying out the following five steps. (This procedure is also illustrated in Fig.\ref{figure:link}.) 

\begin{enumerate}
\item Invert the $6\times 6$ matrix $C^{M\pr{\physg}\veryshortarrow \physg}$ to obtain $c^{M\pr{\physg}\veryshortarrow \physg}$.
\item Using $c^{M\pr{\physg}\veryshortarrow \physg}$ and \eqref{eq:cCompVoigt} determine $c^{M\pr{\physg}\veryshortarrow \physg}_{\pr{(i,j),(k,l)}}$.
\item Using $c^{M\pr{\physg}\veryshortarrow \physg}_{\pr{(i,j),(k,l)}}$ and  \eqref{eq:cconversion} determine  $c_{\pr{(i,j),(k,l)}}^{M\pr{\physg}\veryshortarrow \physe}$. In particular, when $\physg\ag{\mathcal{x}}=\fp\ag{\mathcal{x}}$ equation \ref{eq:cconversion} reads
\begin{equation}
c_{\pr{(i,j),(k,l)}}^{M\pr{\physf_{\varphi}}\veryshortarrow \physe}
= \sum_{p,q,r,s \incylinder} Q_{\cdot p \cdot i}\ag{\varphi} Q_{\cdot q \cdot j}\ag{\varphi} Q_{\cdot r \cdot k}\ag{\varphi} Q_{\cdot s\cdot l}\ag{\varphi}  c_{\pr{(p,q),(r,s)}}^{M\pr{\physf_{\varphi}}\veryshortarrow \physf_{\varphi}},
\label{eq:ftoeconversion}
\end{equation}
where, as defined in \eqref{eq:QDef}, $Q_{\cdot \cdot}\ag{\varphi}$ maps $\physe\ag{\mathcal{x}}$ to $\fp\ag{\mathcal{x}}$. Alternatively, when $\physg\ag{\mathcal{x}}=\physx$ equation \ref{eq:cconversion} takes the form 
\begin{equation}
 c_{\pr{(i,j),(k,l)}}^{M\pr{\physx}\veryshortarrow \physe}
 = \sum_{p,q,r,s \incylinder} P_{\cdot p \cdot i}\ag{X} P_{\cdot q \cdot j}\ag{X} P_{\cdot r \cdot k}\ag{X} P_{\cdot s\cdot l}\ag{X}  c_{\pr{(p,q),(r,s)}}^{M\pr{\physx}\veryshortarrow \physx}
 \footnote{
  For the material models we consider  in this work (\S\ref{subsec:particularization})
 the sum on the right hand side of this  equation comes out to be independent of  $X$ despite the presence of $P_{\cdot i\cdot j}\ag{X}$ in it. For that reason we write $c_{\pr{(i,j),(k,l)}}^{M\pr{\physx}\veryshortarrow \physe}$ on the left hand side instead of $c_{\pr{(i,j),(k,l)}}^{M\pr{\physx}\veryshortarrow \physe}\ag{X}$.},
\label{eq:xtoeconversion}
\end{equation}
where 
\begin{equation}
P_{\dt \dt}\ag{X}
=
\left(\begin{array}{ccc}
\cos\ag{\breve{\theta}\ag{X}} & -\sin\ag{\breve{\theta}\ag{X}}  & 0 \\ [3 pt]
 \sin\ag{\breve{\theta}\ag{X}} &\cos\ag{\breve{\theta}\ag{X}} & 0 \\ [3 pt]
 0 & 0 & 1
\end{array}
\right)
\label{eq:Pdef}
\end{equation}
maps $\physe\ag{\mathcal{x}}$ to $\physx$. 
\item Using $c_{\pr{(i,j),(k,l)}}^{M\pr{\physg}\veryshortarrow \physe}$ and  \eqref{eq:cCompVoigt} determine $c^{M\pr{\physg}\veryshortarrow \physe}$. 
\item Invert the $6\times 6$ matrix $c^{M\pr{\physg}\veryshortarrow \physe}$ to get $C^{M\pr{\physg}\veryshortarrow \physe}$.
\end{enumerate}

\section{Proofs}
\subsection{If there exist three or more $r_i\in (r_{n-1},r_n)$ such that $\tau_{n\cdot 1 \cdot 2}\ag{S_{r_i}}=0$ then $K_{n\cdot i}$, $i=1,\ldots,4$ and $\mu_{n\cdot i}$, $i=1,2$, all vanish.}
\label{sec:step6}

Recall that $r_{n-1}$, $r_n$ are the inner and outer radii of the $n^{\rm th}$ layer in our spicule model. From \eqref{eq:tau12} and \eqref{eq:def:kappa} it follows that  
\begin{equation}
    \tau_{n\cdot 1 \cdot 2}\ag{S_r}=0\Leftrightarrow f_n[r]=0,
\label{eq:tau12fnequivalence}
\end{equation}
where, recall that, $S_r$ is a cylindrical surface of radius $r$, and the function $f_n$ is defined in \eqref{eq:def:fn}. 

In our particularization of the JC model \S\ref{subsec:PaJCModel} we stated that the  layers are able to freely slip with respect to each other, with no friction between them. This implies from \eqref{eq:tau12fnequivalence} that  
\begin{subequations}
\begin{align}
    f_{n}\ag{r_{n-1}}&=0,\\
    f_{n}\ag{r_{n}}&=0.
\end{align}
\label{eq:fnVanishingatBoundaries}
\end{subequations}
That is, the function $f_n$ vanishes at the boundaries of $[r_{n-1},r_n]$. 
In this section we show that if additionally $\tau_{n\cdot 1 \cdot 2}$ vanishes at three or more inner cylindrical surfaces then the constants $K_{n\cdot i}$, $i=1,\ldots,4$, and $\mu_{n\cdot i},~i=1,2$, on which $\tau_{n\cdot 1 \cdot 2}$ and $\tau_{n\cdot 1 \cdot 3}$ depend, all vanish.   

If  $\tau_{n\cdot 1 \cdot 2}$ vanishes at three or more inner cylindrical surfaces with radii $r_1$, $r_2$, and $r_3$ then, as before, it follows from \eqref{eq:tau12fnequivalence} that

\begin{subequations}
\begin{align}
    f_{n}\ag{r_{1}}&=0,\\
    f_{n}\ag{r_{2}}&=0,\\
    f_{n}\ag{r_{3}}&=0.\\
\end{align}
\label{eq:fnVanishingInside}
\end{subequations}

It follows from \eqref{eq:def:fn},  \eqref{eq:fnVanishingInside}, and \eqref{eq:fnVanishingatBoundaries} that 

\begin{equation}
\begin{bmatrix}
e^{\alpha_1 x_1} & e^{\alpha_2 x_1} & e^{\alpha_3 x_1} & e^{\alpha_4 x_1} & e^{\alpha_5 x_1}\\
e^{\alpha_1 x_2} & e^{\alpha_2 x_2} & e^{\alpha_3 x_2} & e^{\alpha_4 x_2} & e^{\alpha_5 x_2}\\
e^{\alpha_1 x_3} & e^{\alpha_2 x_3} & e^{\alpha_3 x_3} & e^{\alpha_4 x_3} & e^{\alpha_5 x_3}\\
e^{\alpha_1 x_4} & e^{\alpha_2 x_4} & e^{\alpha_3 x_4} & e^{\alpha_4 x_4} & e^{\alpha_5 x_4}\\
e^{\alpha_1 x_5} & e^{\alpha_2 x_5} & e^{\alpha_3 x_5} & e^{\alpha_4 x_5} & e^{\alpha_5 x_5}\\
\end{bmatrix}
\begin{bmatrix}
K_{n\cdot 1} \\
K_{n\cdot 2}\\
K_{n\cdot 3}\\
K_{n\cdot 4}\\
\mu_{n\cdot 1}
\end{bmatrix}
=
\begin{bmatrix}
0\\
0\\
0\\
0\\
0
\end{bmatrix},
\label{eq:MKequation}
\end{equation}
where 
\begin{subequations}
\begin{align}
x_i&=\ln{r_i},\quad i=1,\ldots,3, \label{eq:x1to3}\\
x_4&=\ln{r_{n-1}},\label{eq:x4}\\
x_5&=\ln{r_n},\label{eq:x5}
\end{align}
\label{eq:xidefs}
\end{subequations}
and
\begin{subequations}
\begin{align}
\alpha_i&=m_{n\cdot i}-1,\quad i=1,\ldots,4,\\
\alpha_5&=1.    
\end{align}
\label{eq:alphadefs}
\end{subequations}

It can be shown through application of the Rolle's Theorem that when the $\alpha_i$ and $x_i$ in the $\mathcal{M}_{5 \times 5}\pr{\mathbb{R}}$ matrix in \eqref{eq:MKequation} are distinct then that matrix is non-singular~\cite{polya2004problems}. 
The $x_i,~i=1,\ldots,5$ and $\alpha_i,~i=1,\ldots,5$ in \eqref{eq:MKequation} are indeed distinct. We elaborate on these facts in the following two paragraphs.

Recall that in the JC model $r_n>0$ for all $n$, and, without loss of generality, we can take  $r_{n-1}<r_1<r_2<r_3<r_n$. Therefore, it follows from \eqref{eq:xidefs} that the $x_i$, $i=1,\ldots,5$, in \eqref{eq:MKequation} are all distinct.

The equations, e.g., \eqref{eq:tau12} and \eqref{eq:def:fn}, which we used to arrive at  \eqref{eq:MKequation}, are part of the JC theory. In \S\ref{sec:admissibility} we stated that for the JC theory to be applicable to a layer the elastic constants of that layer's helically orthotropic material should be such that their corresponding $m_{n \cdot i}$, $i=1,2$, satisfy the $m$-conditions listed in \eqref{eq:mConditions}. It follows as a consequence of $m_{n\cdot 1}$, $m_{n\cdot 2}$ satisfying the m-conditions and \eqref{eq:alphadefs} that all the $\alpha_i,~i=1,\ldots,5$ are distinct. 

Since, $x_i,~i=1,\ldots,5$ and $\alpha_i,~i=1,\ldots,5$ are distinct the $\mathcal{M}_{5 \times 5}\pr{\mathbb{R}}$ matrix in \eqref{eq:MKequation}  is non-singular and hence,
\begin{subequations}
\begin{align}
K_{n\cdot i}&=0, \quad i=1,\ldots,4, \label{eq:KniZero}\\
\intertext{and}
\mu_{n\cdot 1}&=0\label{eq:muniZero}.
\end{align}
\end{subequations}

It follows from \eqref{eq:tau13} and \eqref{eq:def:kappa} that
\begin{equation}
    \tau_{n\cdot 1 \cdot 3}\ag{S_r}=0\Leftrightarrow h_n\ag{r}=0,
\label{eq:tau13gnequivalence}
\end{equation}
where the function $h_n$ is defined in \eqref{eq:def:hn}.
It also follows from the condition that the layers are able to freely  slip with respect to each other, without any friction, and \eqref{eq:tau13gnequivalence} that 
\begin{align}
    h_{n}\ag{r_{n}}&=0.
\label{eq:gnVanishingatBoundaries}
\end{align}

From equation \ref{eq:KniZero}, equation \ref{eq:gnVanishingatBoundaries}, and  equation \ref{eq:def:hn} we get that $\mu_{n\cdot2}r_n=0$, from which it follows that 
\begin{align}
\mu_{n\cdot2}=0.
\label{eq:mu2Zero}
\end{align}

In summary, we have shown that when there exits three or more interior cylindrical surfaces in the $n^{\rm th}$ layer where $\tau_{n\cdot 1\cdot 2}$ vanishes  then $K_{n\cdot i}$, $i=1,\ldots,4$ and $\mu_{n\cdot i}$, $i=1,2$, all vanish.

\subsection{ The matrix $B_{\cdot \cdot}\ag{\Ce[n]}$ is singular iff $m_{n\cdot 1}=2 \lor m_{n\cdot 2}=2$.}
\label{sec:Bdeterminant}
It follows from \eqref{eqns:mnis} that $m_{n\cdot i}$, $i=1,\ldots, 4$, are roots of the equation

\begin{equation}
d\ag{\Ce[n]}m^4+h\ag{\Ce[n]}m^2+l\ag{\Ce[n]}=0,
\label{eq:charEq2}
\end{equation}
where $d\ag{\cdot}, h\ag{\cdot}$, and $l\ag{\cdot}$ are defined in \eqref{eqns:anbncn}.

Solving for 
$\BetaC[s][2][2]$, 
$\BetaC[s][1][4]$, and
$\BetaC[s][1][1]$, from \eqref{eq:def:d}
\eqref{eq:def:h}, and \eqref{eq:def:l}, respectively, we get that
\begin{subequations}
\begin{align}
\BetaC[s][2][2]&=\frac{d\ag{s}+\p\pr{\BetaC[s][2][4], 2}}{\BetaC[s][4][4]} \label{eq:Beta22},\\
\begin{split}
\BetaC[s][1][4]&=-\BetaC[s][2][4]\pm \p(h \ag{s}+\BetaC[s][1][1]\BetaC[s][4][4]+2\BetaC[s][1][2]\BetaC[s][4][4]\\
&\qquad +\BetaC[s][2][2]\BetaC[s][5][5]-2\BetaC[s][2][4]\BetaC[s][5][6]+\BetaC[s][4][4]\BetaC[s][6][6], 1/2)\label{eq:Beta14}
\end{split},\\
\BetaC[s][1][1]&=\frac{l\ag{s}+\p\pr{\BetaC[s][5][6],2}}{\BetaC[s][5][5]}-(2\BetaC[s][1][2]+\BetaC[s][2][2]+\BetaC[s][6][6])\label{eq:Beta11}.
\end{align}
\end{subequations}

From \eqref{eq:Bcomponents} we have the determinant of $B_{\cdot\cdot}\ag{s}$, 
\begin{multline}
\label{eq:detBform1}
\text{det}\pr{B_{\cdot\cdot}\ag{s}}=-((2\BetaC[s][1][4]+6\BetaC[s][2][4]-\BetaC[s][5][6])(2\BetaC[s][1][4]-2\BetaC[s][2][4]+\BetaC[s][5][6]))\\
+(4\BetaC[s][4][4]-\BetaC[s][5][5])(\BetaC[s][1][1]+2\BetaC[s][1][2]-3\BetaC[s][2][2]+\BetaC[s][6][6]).
\end{multline}
Replacing $\BetaC[s][2][2]$, $\BetaC[s][1][4]$, and $\BetaC[s][1][1]$  in \eqref{eq:detBform1} with, respectively, the right hand sides of \eqref{eq:Beta22}, \eqref{eq:Beta14}, and \eqref{eq:Beta11}, and simplifying we get that
\begin{equation}
\text{det}\pr{B_{\cdot\cdot}\ag{s}}=-(16d\ag{s}+4h\ag{s}+l\ag{s}).
\label{eq:detBForm2}
\end{equation}

If $B_{\cdot\cdot}\ag{\Ce[n]}$ is singular, i.e., its determinant is naught, then it follows from \eqref{eq:detBForm2} that  \begin{equation}
    16d\ag{\Ce[n]}+4 h\ag{\Ce[n]}+l\ag{\Ce[n]}
\label{eq:16d4hl}
\end{equation} is naught as well, which implies that the real number $2$  is a root of \eqref{eq:charEq2}.
Without loss of generality, we take  $m_{n\cdot 3}$, $m_{n\cdot 4}$ to be non-positive.
Thus, we have that when  $B_{\cdot \cdot}\ag{\Ce[n]}$ is singular then $m_{n\cdot 1}=2\lor m_{n\cdot 2}=2$. 

Say $m_{n\cdot 1} =2 \lor m_{n\cdot 2}=2$ then it follows from \eqref{eq:charEq2} that the expression \eqref{eq:16d4hl} again vanishes, which in conjunction with \eqref{eq:detBForm2} implies that $\text{det}\pr{B_{\cdot \cdot}\ag{\Ce[n]}}=0$, i.e., that $B_{\cdot \cdot}\ag{\Ce[n]}$ is singular.

In summary, we have that $m_{n\cdot 1} =2\lor m_{n\cdot 2}=2$ iff $B_{\cdot \cdot}\ag{\Ce[n]}$ is singular.

\bibliographystyle{elsarticle-harv}
\bibliography{RefsBending}

\end{document}